\documentclass[lettersize,journal]{IEEEtran}
\usepackage{amsmath,amsfonts}
\usepackage{algorithmic}
\usepackage{algorithm}
\usepackage{array}
\usepackage{textcomp}
\usepackage{stfloats}
\usepackage{url}
\usepackage{verbatim}
\usepackage{graphicx}
\usepackage{makecell}
\usepackage{cite}
\usepackage[bookmarks=true,hypertexnames=true,
            colorlinks=true,
            linkcolor=black,
            citecolor=blue,
            urlcolor=blue ]{hyperref}
\usepackage[caption=false,font=footnotesize]{subfig}

\begin{document}

\title{Accelerated Transformer Energization Sequence for Inverter Based Resources in Black-Start Procedures with Active Flux Trajectory Manipulation in the Stationary Reference Frame}

\author{Jiyu~Lee,~\IEEEmembership{Student~Member,~IEEE}, 
        Younggi~Lee,~\IEEEmembership{Member,~IEEE},
        Jonghun~Yun,~\IEEEmembership{Member,~IEEE}, 
        Jaekeun~Lee,~\IEEEmembership{Student~Member,~IEEE},
        Heng~Wu,~\IEEEmembership{Member,~IEEE},
        Jae-Jung~Jung,~\IEEEmembership{Member,~IEEE},
        and Shenghui~Cui,~\IEEEmembership{Member,~IEEE}%

}

\markboth{Journal of Emerging and Selected Topics in Power Electronics}%
{Lee \MakeLowercase{\textit{et al.}}: Ultra-Fast Black Start Method for Grid-Forming Converters}



\maketitle

\begin{abstract}
This paper proposes advanced soft-magnetization techniques to enable ultra-fast and reliable black-start of grid-forming (GFM) converters. 
Conventional hard-magnetization with well-established three-phase voltages during transformer energization induces severe inrush currents due to flux offset, which can damage power semiconductor devices. 
To overcome this drawback, an ultra-fast soft-magnetization method is firstly introduced, leveraging the voltage programmability of the inverter to actively reshape the initial voltage profile and thereby eliminate flux offset of the transformer core.
By suppressing the formation of flux offset itself, the proposed approach prevents magnetic saturation and achieves nominal terminal voltage within a few milliseconds while effectively suppressing inrush current.
However, this method can still trigger surge currents to power semiconductor devices in the presence of an LC filter due to abrupt voltage magnitude and phase transitions. 
To address this issue, an enhanced Archimedean spiral soft-magnetization method is developed, where both voltage magnitude and phase evolve smoothly to simultaneously suppress inrush and surge currents.
Furthermore, residual flux in the transformer core is considered, and a demagnetization sequence using the inverter is validated to ensure reliable start-up.
Experimental results confirm that the proposed methods achieve rapid black-start performance within one fundamental cycle while ensuring safe and stable operation of GFM converters.
\end{abstract}


\section{Introduction}
 \IEEEPARstart{T}{he} grid-forming (GFM) converter, originally introduced for application in microgrids, is gaining attention as it can operate stably even in modern grids where traditional synchronous generators are being replaced by inverter-based resources (IBRs) \cite{GFM_Contol_Basic,rosso_grid-forming_2021,GFM_IBR}.
 Unlike the conventional grid-following (GFL) converter, which relies on a phase-locked loop (PLL) to synchronize with an external voltage source, the GFM converter inherently establishes the voltage and frequency reference of the system.
 This capability allows it to function as an independent voltage source, thereby supporting standalone operation in de-energized networks without the presence of any external voltage reference.

\ In this regard, GFM converters are required to provide black-start capability, i.e., the ability to re-energize the system after a blackout without assistance from external voltage sources \cite{burroughs_black_2023,Black_cap_GFM,Black_start_cap2}.
 During black-start, abrupt voltage injection of balanced and well-established three-phase voltages with full magnitude like a synchronous generator can drive the transformer core into severe magnetic saturation by shifting the flux trajectory above the knee point of its nonlinear magnetization curve.
 As a result, the effective magnetizing inductance collapses and severe inrush currents are induced. 
 Unlike synchronous generators—which can withstand transient overcurrents of 5–7 p.u.—IBRs are constrained by the safe-operating area of their power semiconductors, which in turn limits their overcurrent capability and permissible duration.
 Therefore, to prevent converter failure, extensive research has focused on mitigating transformer inrush currents during black-start.

\ Among hardware-based approaches, the transient inrush current was mitigated in \cite{PIR} by increasing the source feeding impedance during the energization process using a pre-insertion resistor (PIR). 
 However, this method requires modifications to the hardware structure, resulting in a bulkier overall system and making it cost-ineffective. 
 Software-based approaches have also been proposed, including the controlled switching (CS) method \cite{CS_Method} and prefluxing with virtual resistor \cite{Prefluxing_VR}.
 In \cite{CS_Method}, the circuit breaker is closed at an optimal instant aligned with the residual flux of the transformer, allowing energization at the rated voltage. However, this method requires complex pre-calculations and demands high accuracy in measuring the transformer’s flux.
 Moreover, different from the solid-state switches, the stochastic delay time of closing action of large-scale mechanical circuit breakers makes application of CS more challenging in practice.
 In \cite{Prefluxing_VR}, transformer inrush is mitigated by pre-fluxing the core through inverter switching states, followed by energization at 180° phase to minimize flux offset. However, this method models each phase independently and does not explicitly consider inter-phase flux coupling in three-phase transformers.
 
 \ The magnetization methods proposed in \cite{PIR,CS_Method,Prefluxing_VR} assume that the three-phase voltages are well-established at the rated magnitude before connected to the transformer for magnetization.
 This is inevitably true when synchronous generators are considered as enabler of black-start. 
 Note that different from the synchronous generators, the output voltage profile of GFM converters are fully programmable in terms of instantaneous magnitude and phase.
 Therefore, black-start of power systems with GFM converter provides additional control degree of freedom.
 As simpler and more widely adopted alternatives, various soft-energization methods by GFM converters gradually ramp up the voltage reference \cite{alassi_performance_2020,NRE, Soft_Energization2}. 
 However, these approaches rely on natural decay of excessive magnetization current, which is very slow in a range of several seconds to dozens of seconds in the case of high-power transformers.
 Therefore, considerable delays on the black-start process is imposed even in purely electronic AC grids.
 Moreover, very slow ramps prolong the time during which protection relays lose the capability to detect faults due to the establishment of voltage being too slow \cite{alassi_performance_2020}.
 To overcome these drawbacks, \cite{Ramp_Time_Estimation} proposes a method to estimate the minimum voltage ramp-up time quantitatively. 
 However, since the approach still relies on the natural decay of core flux, the achievable start-up time remain limited to the order of several seconds.
 
\ This paper proposes a transformer soft-magnetization method that enables fast and reliable black-start by actively suppressing the formation of flux offset itself. 
 By leveraging the voltage profile controllability of the inverter, the proposed method actively adjusts the initial voltage injection profile to eliminate flux offset linked to the transformer core, thereby preventing magnetic saturation.
 Hence, this method enables the terminal voltage to reach its nominal value within just a few milliseconds, while effectively suppressing inrush current. 
 
 Compared to the previous work presented in the conference paper version by the authors in \cite{PEDG}, which introduced the ultra-fast soft-magnetization method, this post-conference journal additionally addresses two critical limitations: surge currents caused by the LC filter and the presence of residual flux in the transformer core.
 While the ultra-fast soft-magnetization method in \cite{PEDG} effectively eliminated flux offset and suppressed inrush current of the transformer, it did not consider the resonance-induced surge current of the LC filter to the power semiconductor devices nor the adverse effects of residual flux remaining after blackout events on the application of the proposed soft-magnetization method.
 In contrast, this paper presents an enhanced soft-magnetization strategy that profiles the applied voltage using an Archimedean spiral trajectory to mitigate surge currents to power semiconductor devices.
 Furthermore, the residual core flux left after the blackout is taken into account, and it is verified that a reliable black-start can be achieved through a subsequent demagnetization process driven by the inverter.

\ The remainder of this paper is organized as follows. Section~\ref{Section2} describes the control strategy of the GFM converter and analyzes the cause of flux offset by transforming the three-phase voltage and flux into vectors in the stationary reference frame.
 Section~\ref{Proposed} presents a vector-based approach for designing a voltage profile that eliminates flux offset \cite{PEDG} and proposes an improved black-start voltage profile that accounts for surge currents induced by LC filters.
 Section~\ref{Experiments} validates the proposed method through experiments and further investigates the impact of residual magnetic flux. To address the residual flux issues, a pre-demagnetization sequence is introduced and experimentally verified.
 Lastly, Section~\ref{conclusion} concludes this paper.
\section{Analysis on the Applied Voltage and the Flux Linkage of the Transformer during Black-Start}
\label{Section2}

During the black-start of GFM converters, the applied voltage plays a crucial role in determining the transient behavior of the transformer core flux linkage. 
In this section, both the control strategy of GFM converters and the theoretical analysis of transformer flux dynamics are presented.

\begin{figure}[t]
    \centering
    \includegraphics[width= 0.95\linewidth]{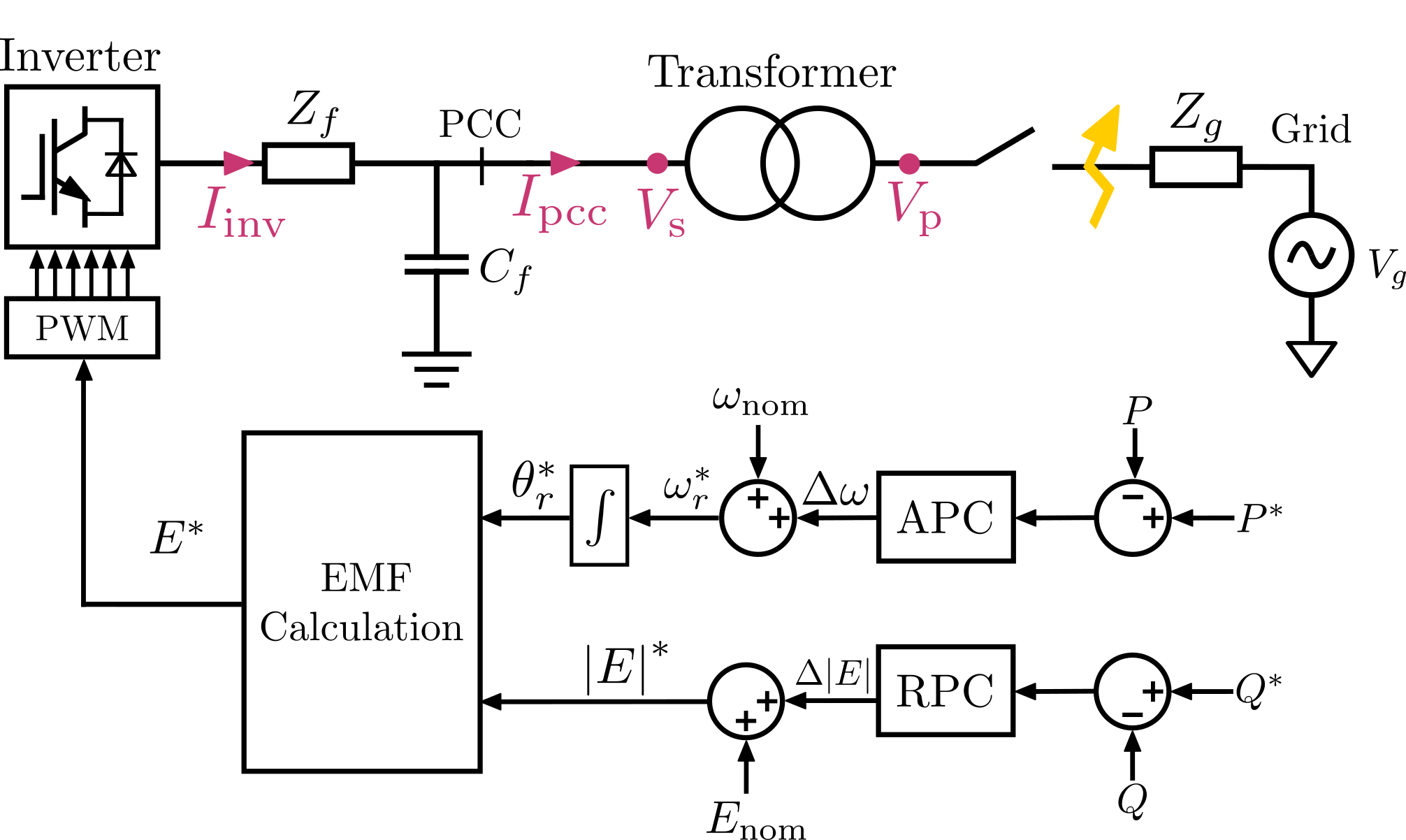}
    \caption{Control block diagram of a GFM converter.}
    \label{GFM_controller}
\end{figure}

\subsection{Control Strategy of the GFM Converters}
  The control structure of the GFM converter addressed in this paper is shown in Fig.~\ref{GFM_controller}. 
  The GFM converter with an LC filter is connected to the grid via a transformer at the point of common coupling (PCC). 
 The active power controller (APC) generates the angular frequency reference of the virtual synchronous generator based on the error between the active power reference and the measured active power output. The APC can be implemented using various schemes such as droop control \cite{PSC} or synchronous power control (SPC) \cite{SPC}, which emulates virtual inertia.
 The voltage magnitude reference of the converter is generated by the reactive power controller (RPC). In this paper, the RPC is implemented using a droop relationship between the voltage magnitude and the error between the reactive power reference and the measured reactive power. Also, a simplified single-loop control structure is adopted, wherein no inner-loop current controller is employed, thereby enabling a more direct modulation of the converter voltage reference.
 When a blackout occurs and the converter becomes islanded from the main grid, the electromotive force (EMF) reference formed by the APC and RPC serves as the new voltage and frequency reference for the system.
 With APC and RPC, the output voltage of the GFM converter is well-established at rated magnitude and frequency before connecting to the transformer for magnetization. 
 Since the voltage formed by the GFM converter is applied through the transformer, the energization process inherently involves the magnetization of the transformer core prior to supplying power to the remainder of the grid. This initial magnetization phase plays a critical role in determining the transient behavior of the system and should be carefully controlled to prevent core saturation and the resulting inrush current.

\begin{figure}[!t]
    \centering
    \subfloat[]{\includegraphics[width=0.45\linewidth]{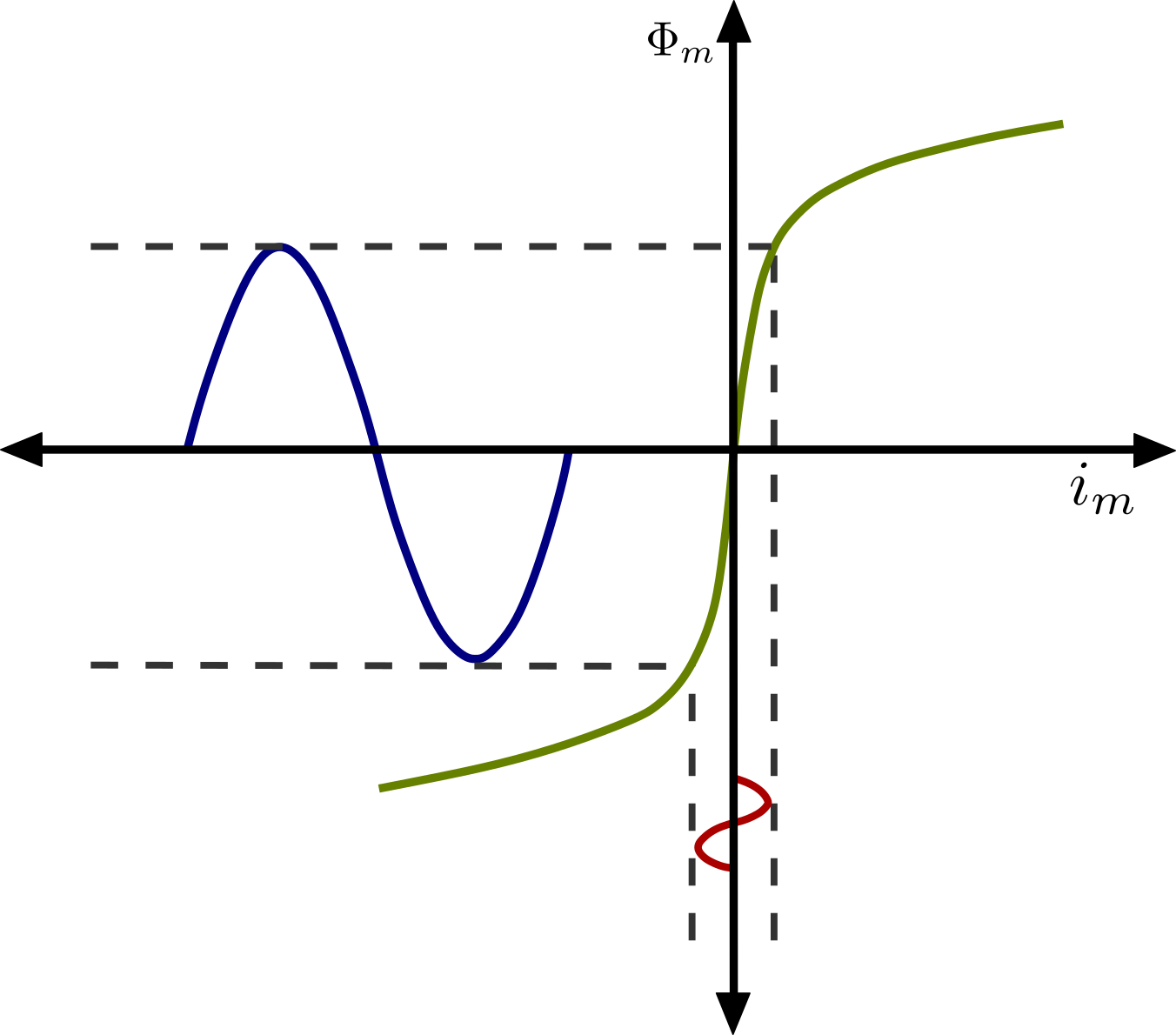}
      \label{Fig2_a}}
    \subfloat[]{\includegraphics[width=0.45\linewidth]{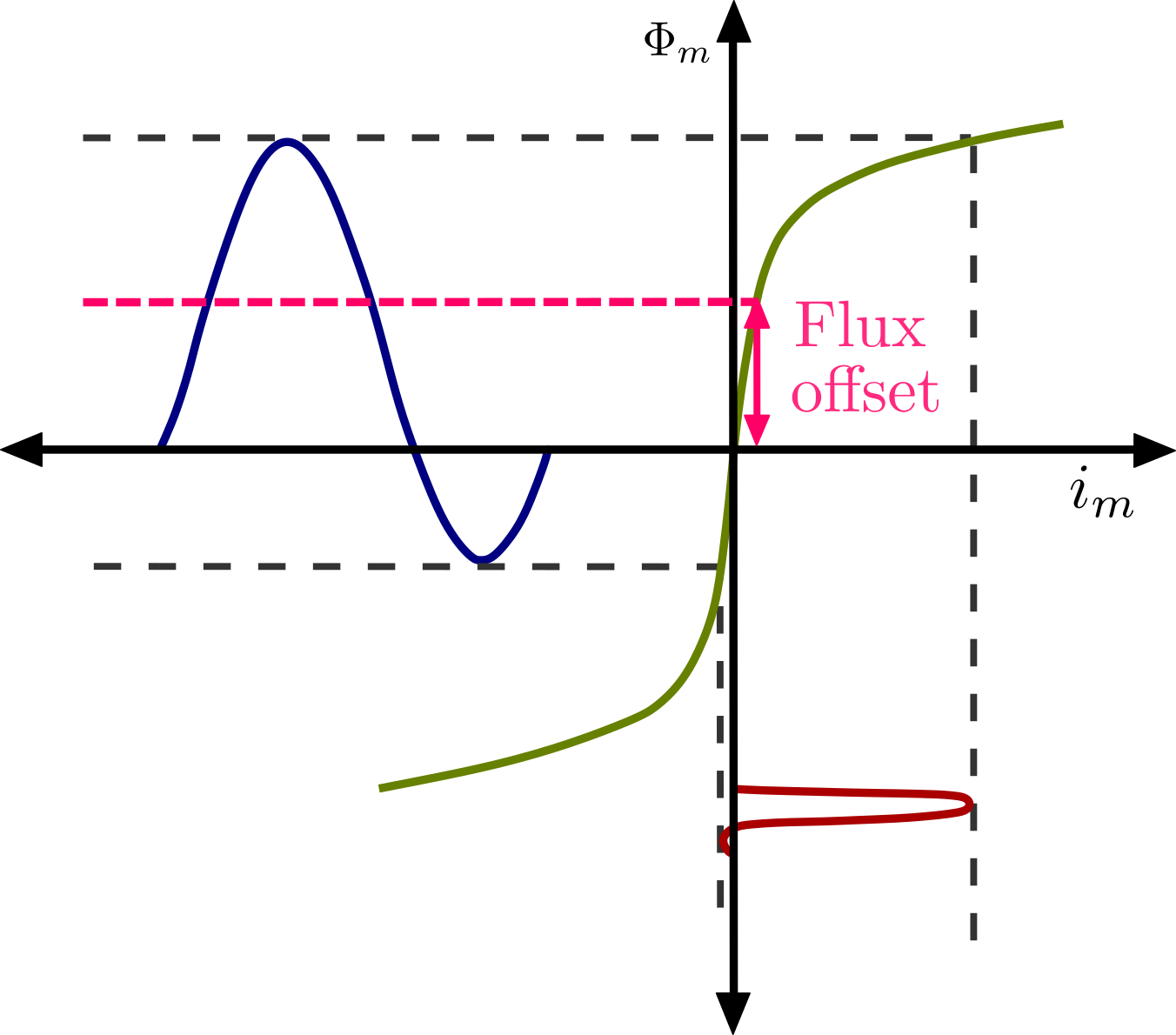}
      \label{Fig2_b}}
    \caption{Transformer flux linkage and magnetizing current: (a) normal operation, (b) magnetic core saturation.}
    \label{Fig2}
\end{figure}

\subsection{Theoretical Analysis of Flux Bias and Transformer Core Saturation}
Fig.~\ref{Fig2} illustrates the conceptual relationship between the transformer core flux and the resulting magnetizing current.
The blue sinusoid corresponds to the flux linkage induced by the applied AC voltage, whereas the red sinusoid corresponds to the magnetizing current.
Under normal operating conditions, when an AC voltage is applied to a transformer with an open secondary winding, only a small magnetizing current flows, as shown in Fig.~\ref{Fig2}(a).
However, if the transformer is excited with an offset in the flux trajectory, such as in Fig.~\ref{Fig2}(b), the operating point can exceed the knee point of the core’s nonlinear magnetization curve.
This leads to a dramatic reduction in the effective magnetizing inductance, causing a substantial inrush current even under the same applied AC voltage.

\ If a balanced three-phase voltage with full magnitude is applied to the transformer instantaneously during black-start, a flux offset is inevitably induced in the transformer core, driving it into magnetic saturation. 
This flux offset arises from the fact that the linked flux is the continuous time integral of the applied voltage. To intuitively analyze this phenomenon, the three-phase voltages and corresponding fluxes can be represented in the stationary reference frame  using Clarke's transformation as follows.

\begin{figure}[!t]
    \centering
    \subfloat[]{\includegraphics[width=0.45\linewidth]{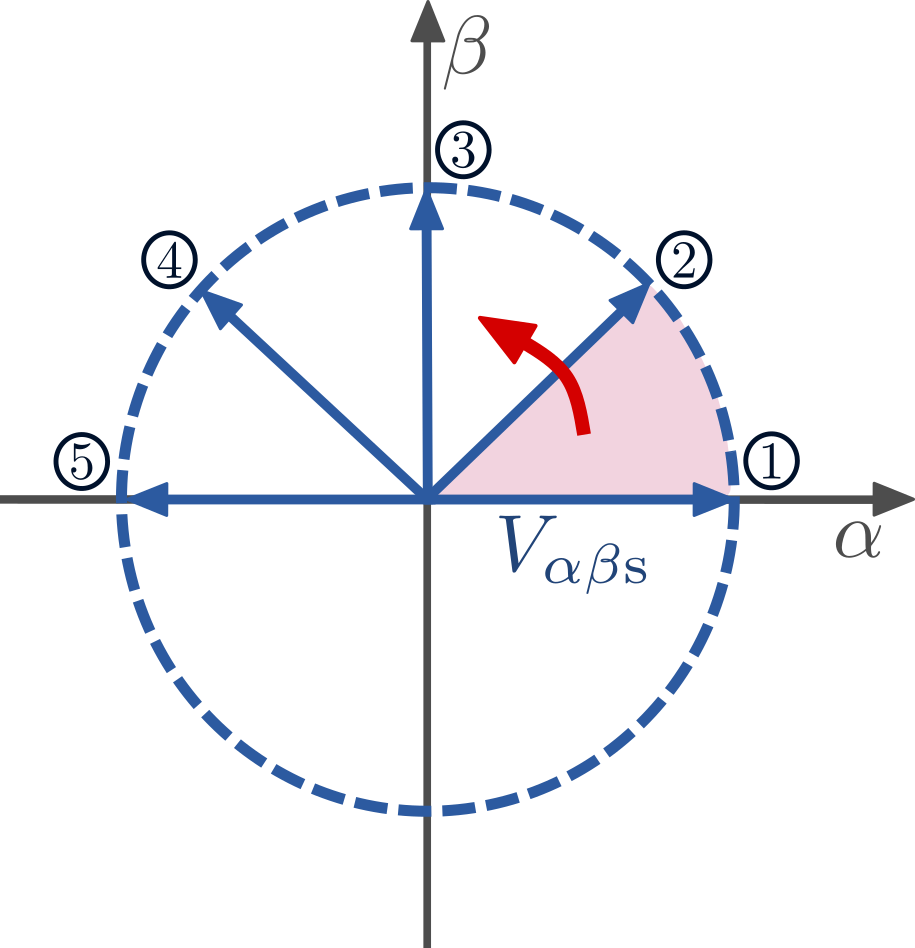}
      \label{Fig3_a}}
    \subfloat[]{\includegraphics[width=0.45\linewidth]{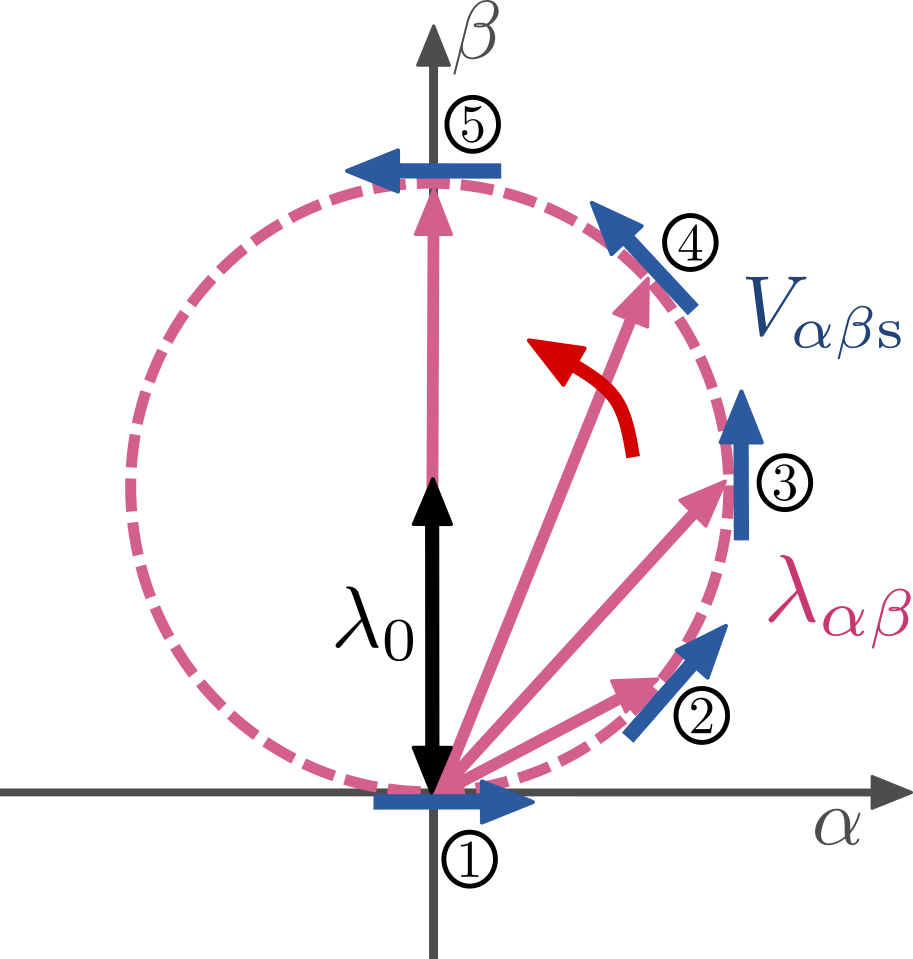}
      \label{Fig3_b}}
    \caption{Vector diagram of instantaneous voltage injection in the stationary reference frame: (a) applied voltage to the transformer, (b) magnetic flux linked to the transformer.}
    \label{Fig3}
\end{figure}

\begin{equation}
  \label{Eq1}
   V_{\text{abcs}}
   =
    \begin{bmatrix}
    v_{\text{as}}(t)\\
    v_{\text{bs}}(t)\\
    v_{\text{cs}}(t)
  \end{bmatrix}
  =
    \begin{bmatrix}
    \hat{V} \cos(\omega_{0} t)\\
    \hat{V} \cos\left(\omega_{0} t - \frac{2}{3}\pi\right) \\
    \hat{V} \cos\left(\omega_{0} t + \frac{2}{3}\pi\right)
    \end{bmatrix}.
\end{equation}

\begin{equation}
  \label{Eq2}
  V_{\alpha \beta \text{s}}
  =
  \begin{bmatrix}
    v_{\alpha \text{s}}(t)\\
    v_{\beta \text{s}}(t)
  \end{bmatrix}
  = 
  \begin{bmatrix}
    1 & -\frac{1}{2} & -\frac{1}{2}\\
    0 & \frac{\sqrt{3}}{2} & -\frac{\sqrt{3}}{2}\\
  \end{bmatrix}
  \cdot
  V_{\text{abcs}}
  =
    \begin{bmatrix}
    \hat{V} \cos(\omega_{0} t)\\
    \hat{V} \sin(\omega_{0} t)
    \end{bmatrix}.
\end{equation}

Neglecting the voltage drop across the winding resistance, the applied voltage corresponds to the time derivative of the magnetic flux linkage \cite{Flux_Volt_JH1,Flux_Volt_JH2,Flux_Volt_JH3}, and the following equation holds:

\begin{equation}
  \label{Eq3}
  \boldsymbol{\lambda}_{\alpha\beta} = \int_{0}^{t} \mathbf{V}_{\alpha\beta\mathrm{s}}\, d\tau = 
  \begin{bmatrix}
    \lambda_{\alpha}(t)\\
    \lambda_{\beta}(t)
  \end{bmatrix}
  =
  \begin{bmatrix}
    \frac{\hat{V}}{\omega_0} \sin(\omega_{0} t)\\
    \frac{\hat{V}}{\omega_0} - \frac{\hat{V}}{\omega_0} \cos(\omega_{0} t)
  \end{bmatrix}.
\end{equation}

Here, $\boldsymbol{\lambda}$ and $\mathbf{V}$ denote the flux linkage and the voltage vector, respectively, while the subscripts and '$s$' refer to the secondary side of the transformer. The term $\hat{V}$ denotes the rated voltage magnitude, and $\omega_0$ represents the rated angular frequency.
As illustrated in \eqref{Eq3}, when three-phase rated voltages are applied in a stepwise manner as in \eqref{Eq1}, a flux offset is introduced along the $\beta$-axis. The resulting flux offset can be expressed as:

\begin{equation}
	\label{Eq4}
    \begin{split}
    \lambda_{0} &= \frac{\hat{V}}{\omega_{0}} .
    \end{split}
\end{equation}

As a result, this flux offset drives the transformer core into deep saturation, thereby inducing significant inrush current during the black-start process.
This is attributed to the fact that grid interface transformers are normally configured in three-phase three-limb structure, which therefore does not necessitate the consideration of zero-sequence component of the flux linkage.

\ The formation of flux offset, as derived analytically, can be further interpreted through vector analysis in the stationary reference frame. Fig.~\ref{Fig3} presents the applied voltage vector and the resulting flux trajectory across the transformer in the $\alpha$–$\beta$ plane.
As the voltage vector continuously rotates from 1 to 5 as shown in Fig.~\ref{Fig3}(a), the integral trajectory of the voltage, corresponding to the flux, traces a biased circular trajectory centered along the $\beta$-axis, as shown in Fig.~\ref{Fig3}(b).
As derived in \eqref{Eq4}, the magnitude of the flux offset is equal to $\lambda_0$, which corresponds to the radius of the resulting circular flux trajectory.
It can be noticed in Fig.~\ref{Fig3} that since the applied voltage is the time derivative of the flux linkage, the tangent vector of the circular flux trajectory aligns with the direction of the injected voltage vector, as indicated by the blue arrow in Fig.~\ref{Fig3}(b).

\section{Proposed Soft-Magnetization Method}
\label{Proposed}

Building on the analysis in Section~\ref{Section2}, this section introduces two soft-magnetization strategies for GFM converters.
By actively profiling the applied voltage, these methods eliminate flux offset and suppress inrush current, enabling reliable transformer energization during black-start.

\subsection{Ultra-Fast Soft-Magnetization Method}
\label{PEDG}
\begin{figure}[!t]
    \centering
    \subfloat[]{\includegraphics[width=0.45\linewidth]{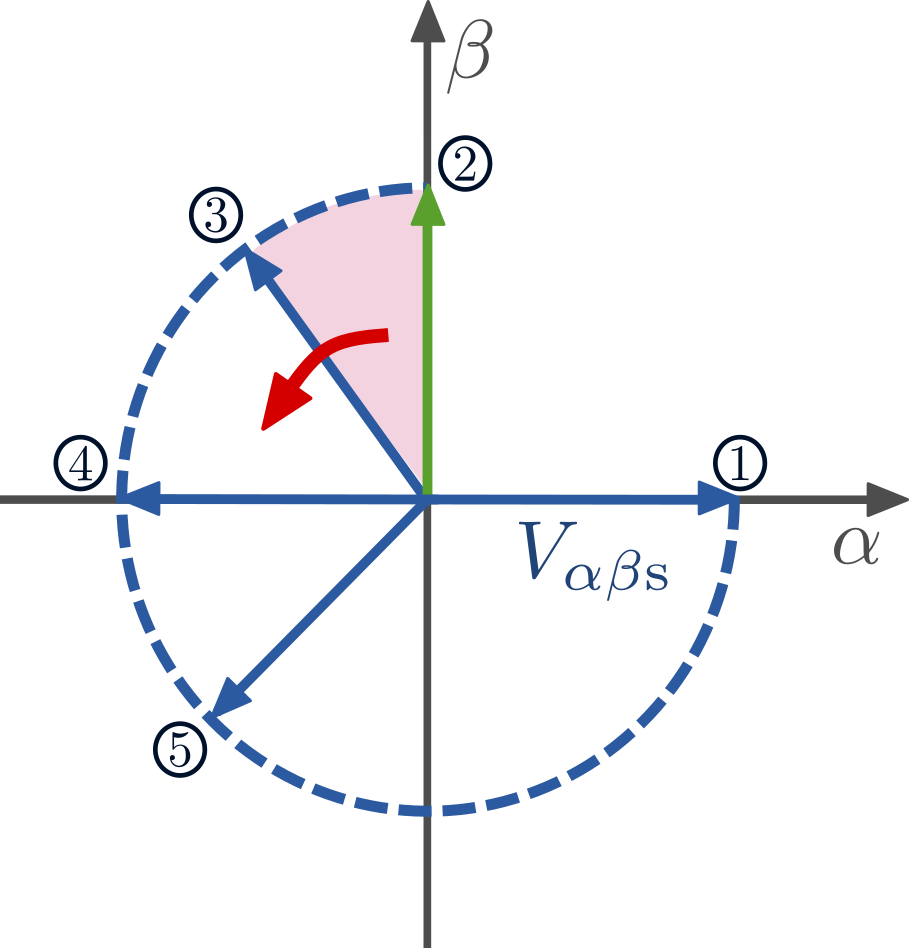}
      \label{Fig4_a}}
    \subfloat[]{\includegraphics[width=0.45\linewidth]{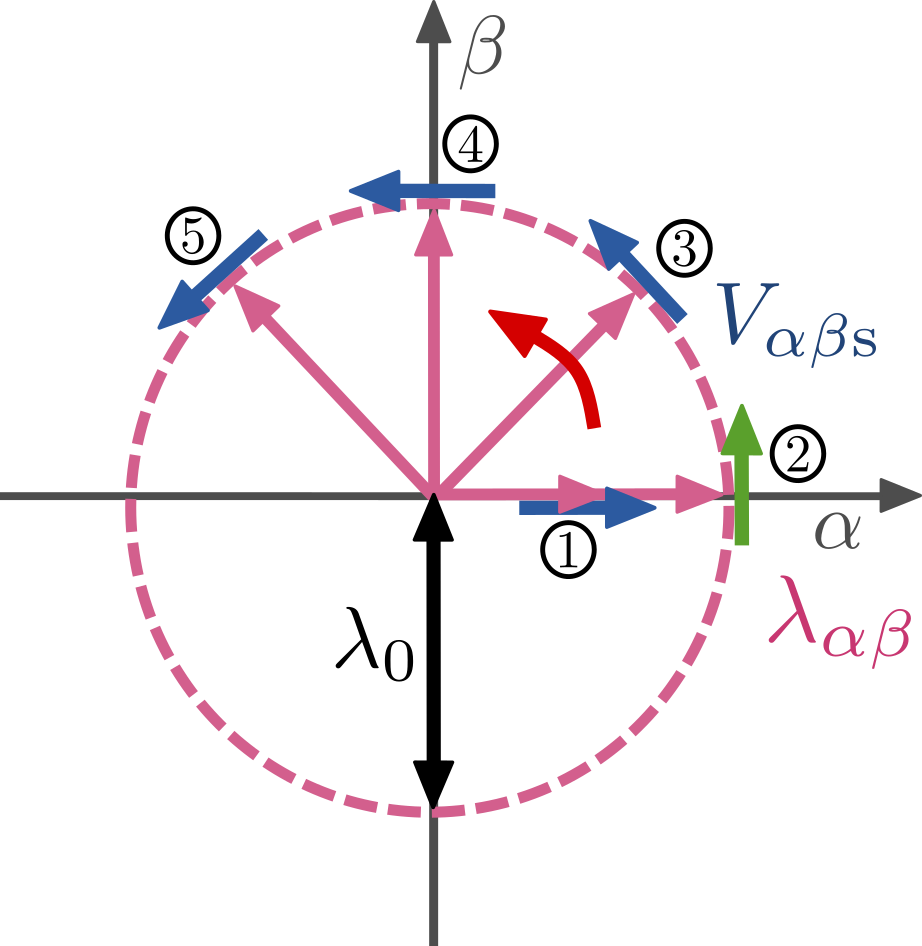}
      \label{Fig4_b}}
    \caption{Vector diagram of ultra-fast soft-magnetization method in the stationary reference frame: (a) applied voltage to the transformer, (b) magnetic flux linked to the transformer.}
    \label{Fig4}
\end{figure}

The proposed method aims to eliminate the flux offset caused by the continuous integration of the voltage vector \cite{PEDG}.
Leveraging the GFM converter’s programmable direct voltage-forming capability, the applied voltage can be actively profiled to prevent the formation of flux offset in the trajectory.
 By injecting a constant voltage vector into the transformer for a certain duration before the voltage vector begins to rotate, the initial flux vector is increased to $\lambda_0$,  effectively centering the flux linkage trajectory at the origin. 
 The injected voltage vector and the corresponding flux trajectory under the proposed method are shown in Fig.~\ref{Fig4}.

 \ First, the voltage vector 1, aligned to $\alpha$-axis with its full magnitude, is applied until the magnitude of the flux vector is increased and the trajectory is shifted to $(\lambda_0,0)$. 
 During this initial period, the applied constant voltage builds up the magnetizing flux of the transformer without introducing any rotational motion. 
 The application duration $T_D$ is determined as follows:

 \begin{equation}
	\label{Eq5}
            T_D = \frac{\lambda_{0}}{\hat{V}} = \frac{1}{\omega_{0}} = \frac{T_0}{2\pi}.
\end{equation}

\ $T_0$ represents the fundamental voltage period. 
Once the flux vector reaches $(\lambda_0,0)$, the voltage vector is shifted to the perpendicular one, namely vector 2, as indicated by the green arrow in Fig.~\ref{Fig4}(b). 
Because the trajectory of the flux is determined by the time integral of the applied voltage, a direction of the voltage vector becomes tangent to the previous flux trajectory. 
Therefore, the flux trajectory changes its direction to become perpendicular, aligned with the perimeter of the newly formed circular path. As the voltage vector starts to rotate after duration $T$, the flux trajectory that reaches the magnitude of $\lambda_0$ begins to trace a new circle centered at the origin. As a result, the flux linkage offset in the transformer core is avoided inherently, successfully preventing core saturation.

Since the phase angle of the voltage vector is determined by the APC of the GFM converter, this method can be implemented easily by initializing the APC output as follows:
 \begin{equation}
 \label{Eq6}
     \theta_{r}^{*}(t) = 
     \begin{cases}
     0,& t < T_D \\
     \frac{\pi}{2},& t = T_D \\
     \frac{\pi}{2} + \int \omega_{r}^{*}\,dt,& t > T_D.
 \end{cases} 
 \end{equation}

   \begin{figure}[t]
    \centering
    \includegraphics[width= 0.95\linewidth]{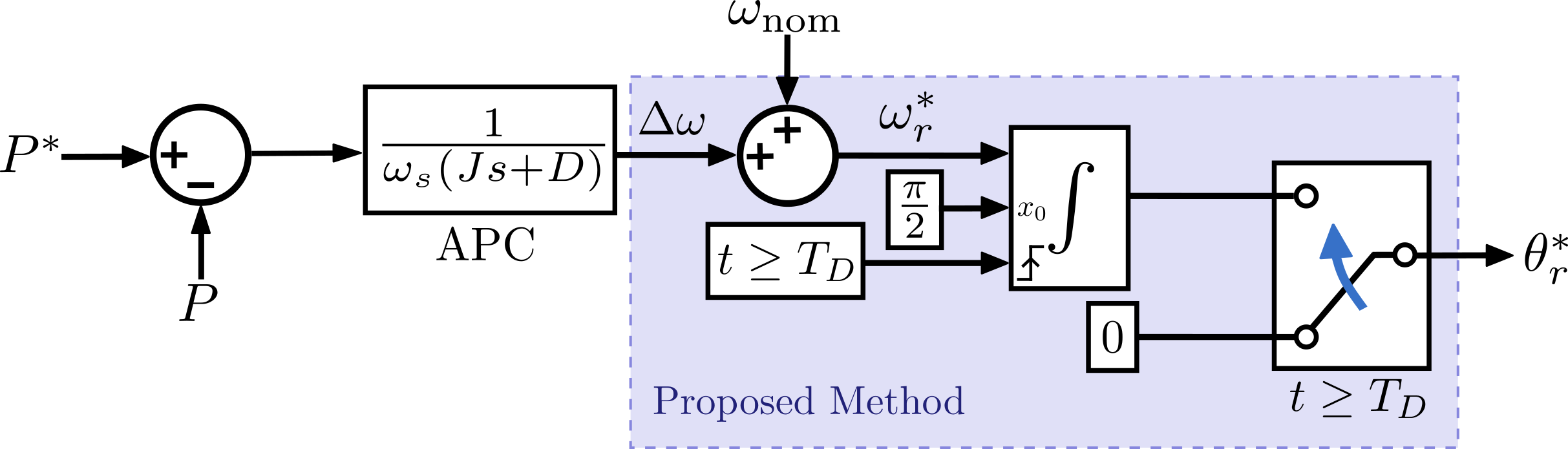}
    \caption{Control block diagram of the proposed ultra-fast soft-magnetization.}
    \label{Proposed_diagram}
\end{figure}

  In (\ref{Eq6}), $\theta_{r}^{*}$ and $\omega_r^{*}$ refer to the phase angle reference and the angular speed reference of the converter, respectively. The control block diagram of the proposed method is presented in Fig.~\ref{Proposed_diagram}. 
  The converter phase angle reference $\theta_{r}^{*}$ is maintained at 0 for $t < T_D$ after start-up, and the integrator state is initialized to $\frac{\pi}{2}$ at $t = T_D$. 
  By simply modifying the initial profile of the injected voltage, the transformer energization can be completed within $\frac{1}{2\pi}$ of the fundamental period, achieving an extremely fast start-up compared to conventional soft-start methods \cite{alassi_performance_2020,NRE, Soft_Energization2}. 

 \ The proposed method can be implemented without any modifications to the controller or limitations on its output, making it highly adaptive to all GFM control strategies. 
  Additionally, the soft-magnetization of the transformer can be implemented with any different arrangement of the flux trajectory, which regulates the trajectory of flux vector actively to avoid DC offset based on this principle. 
  Among various approaches, this method applies a full-magnitude DC voltage to build up the flux by $\lambda_0$ at initial state, making it one of the fastest strategies for transformer magnetization without offset. 
  However, additional considerations involving other practical issues will be taken into account as in the following section.

\subsection{Effects of LC Filter Surge Currents}
 The method presented in Section~\ref{PEDG} enables ultra-fast transformer magnetization by applying a full-magnitude DC voltage.
 However, this approach inherently involves abrupt changes in both the voltage magnitude and phase angle as shown in Fig.~\ref{Fig4}.
 Specifically, the voltage vector is injected with its full amplitude from the very beginning, and the instantaneous phase shift occurs after $T_D$. 
 This, in turn, may excite resonance when an LC filter is present at the converter output.
 Such abrupt excitation pattern can induce high-frequency transients and may pose a threat to the power semiconductor devices or connected loads during the black-start process, even if the transformer is magnetized softly.
 Therefore, although this method can effectively suppress the inrush current caused by magnetic flux saturation, the surge current due to the LC filter resonance might still pose a risk of the power semiconductor device failure in the converter.

\subsection{Enhanced Soft-Magnetization Using Smooth Archimedean Spiral Voltage under LC Filter}

\ To address the surge current issue, this section proposes an enhanced soft-magnetization method that explicitly accounts for the LC filter.
In GFM converter with an LC filter interface, the voltage profile should be designed to suppress flux offset while avoiding abrupt changes in both magnitude and phase.
To this end, the applied voltage vector is shaped according to an Archimedean spiral curve in the stationary reference frame, ensuring that the magnitude and the phase angle vary smoothly and the DC bias of the flux trajectory is avoided throughout the black-start process.
The reference voltage to be applied can be defined as:

\begin{equation}
  \label{Snail_start}
  \begin{bmatrix}
    v_{\alpha \text{s}}(t)\\
    v_{\beta \text{s}}(t)
  \end{bmatrix}
  =
    \begin{bmatrix}
    A\omega t \cos(B \omega t)\\
    A \omega t \sin(B \omega t)
    \end{bmatrix}.
\end{equation}

At the end of the start-up interval $T_A$, the applied voltage and the resulting flux are required to reach their rated values, which leads to the following equation:

\begin{equation}
  \label{Snail1}
  \begin{aligned}
    A \omega T_A \cos{(B \omega T_A)} &= \hat{V}, \\
    A \omega T_A &= \hat{V}
  \end{aligned}
\end{equation}

In addition, the flux magnitude is required to increase by $\lambda_0$ after the duration $T_A$. 
Assuming that the start-up is completed when the voltage vector completes an integer number of rotations, and taking into account that the $\frac{\pi}{2}$ phase difference between the flux and the voltage, the following relation can be derived: 

\begin{equation}
    B \omega T_A = 2 \pi.
\end{equation}

\begin{align}
  \label{Snail2}
\int_{0}^{T_A} A \omega t \cos(B \omega t)\, dt 
&= \left[ A \omega t \frac{1}{B \omega} \sin(B \omega t) \right. \nonumber \\
&\quad \left. + \frac{A}{B \omega^{2}} \cos(B \omega t) \right]_{0}^{T_A} 
= 0, \\
\int_{0}^{T_A} A \omega t \sin(B \omega t)\, dt 
&= \left[ -A \omega t \frac{1}{B \omega} \cos(B \omega t) \right. \nonumber \\
&\quad \left. + \frac{A}{B \omega^{2}} \sin(B \omega t) \right]_{0}^{T_A} 
= -\frac{A}{B}T_A = \lambda_{0}.
\end{align}

By assuming that the voltage vector rotates at the rated angular frequency and by setting the unknown parameters such that the start-up is completed within one rotation, the following expression can be obtained:

\begin{equation}
  \label{Snail3}
  \begin{aligned}
    B &= 1, \\
    T_A &= \frac{2 \pi}{\omega_0} = T_0, \\
    A &= \frac{\hat{V}}{2 \pi}.
  \end{aligned}
\end{equation}

\begin{figure}[!t]
    \centering
    \subfloat[]{\includegraphics[width=0.45\linewidth]{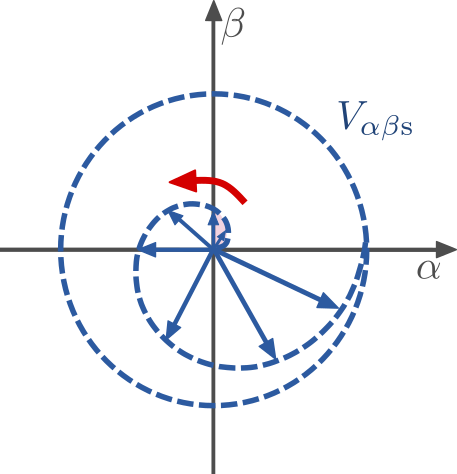}
      \label{Fig6_a}}
    \subfloat[]{\includegraphics[width=0.45\linewidth]{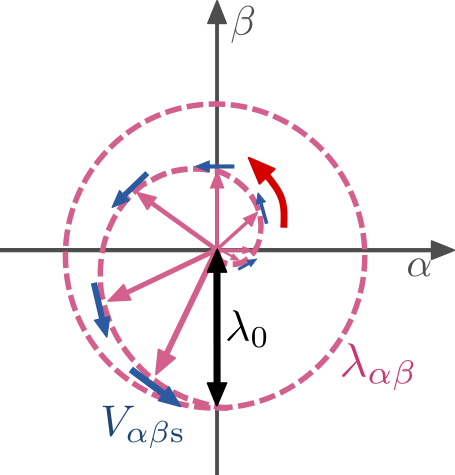}
      \label{Fig6_b}}
    \caption{Vector diagram of enhanced Archimedean spiral soft-magnetization method in the stationary reference frame: (a) applied voltage to the transformer, (b) magnetic flux linked to the transformer.}
    \label{Fig6}
\end{figure}

Therefore, by setting the initial voltage ramp-up slope $A$ as defined above, the start-up time $T_A$ becomes equal to the rated period, and the flux trajectory forms a circle centered at the origin.
By applying the analytically derived slope $A$ during the voltage ramp-up in the black-start process, the voltage and flux vectors in the stationary reference frame can be evaluated as illustrated in Fig.~\ref{Fig6}.
During the start-up period $T_A$, the applied voltage ramps linearly 
from zero to its rated value with slope $A$, while the voltage vector 
completes one full revolution as illustrated in Fig.~\ref{Fig6}(a). 
In this process, the flux trajectory follows an Archimedean spiral, as shown in Fig.~\ref{Fig6}(b),
rising smoothly from zero to the rated magnitude and thereby 
achieving magnetization without introducing any DC offset. 
Since both the magnitude and phase of the voltage vector increase 
smoothly without abrupt transitions, the method can be effectively 
applied to the black-start of GFM converters interfaced  to the transformer through an LC filter.

\begin{figure}[!t]
    \centering
    \includegraphics[width= 0.8\linewidth]{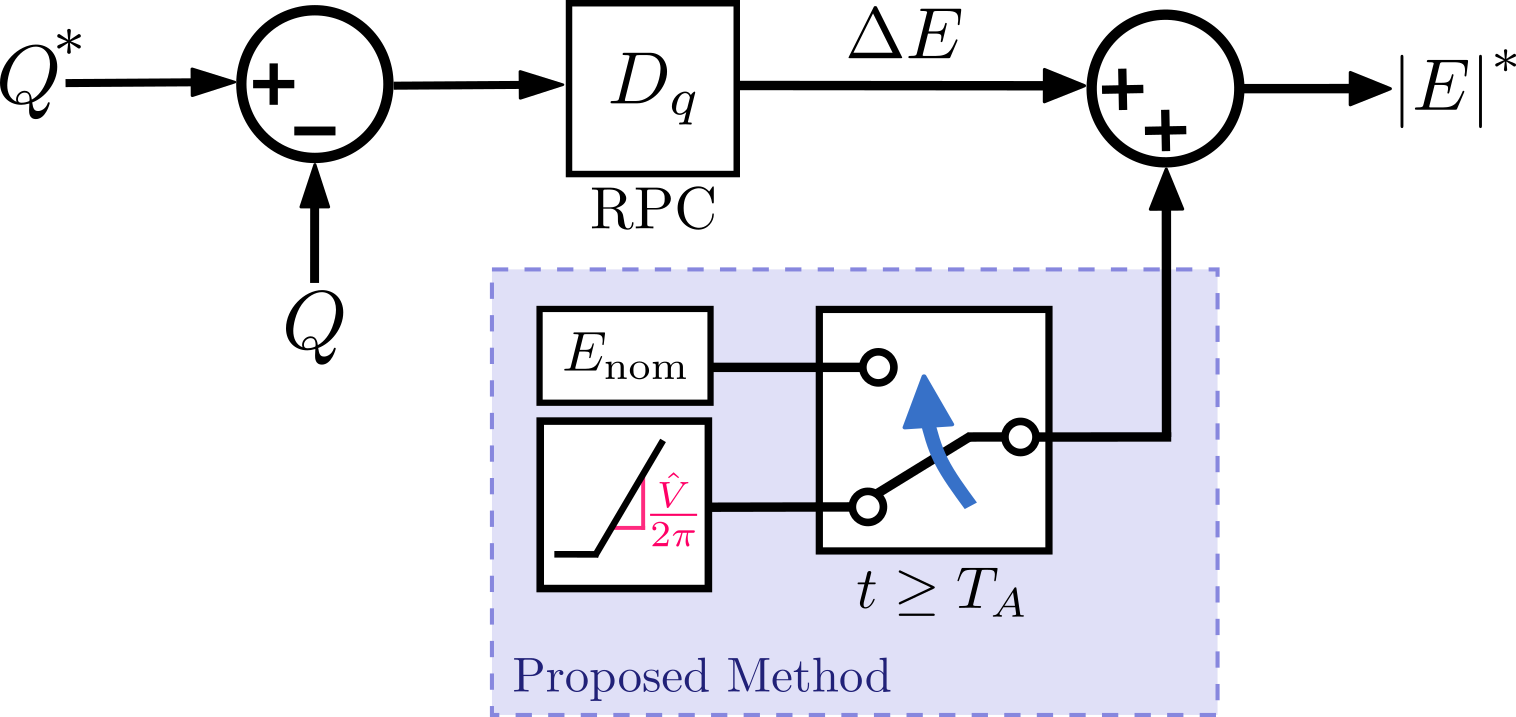}
    \caption{Control block diagram of the enhanced Archimedean spiral soft-magnetization.}
    \label{Proposed_Snail_diagram}
\end{figure}

\ The proposed method controls the voltage magnitude, which can be implemented through the RPC within the GFM control structure. 
Fig.~\ref{Proposed_Snail_diagram} illustrates the block diagram of the proposed Archimedean spiral method. 
The final EMF magnitude reference is obtained by adding a nominal voltage to the output of the reactive power controller. 
During the black-start process, this nominal voltage is ramped from zero to the rated value with a slope of $\tfrac{\hat{V}}{2\pi}$ until the start-up time $T_A = T_0$. 
After $T_A$, the voltage naturally settles at its rated value, ensuring that the transformer is fully magnetized and the start-up process is completed.
Note that the proposed method can be universally applied to all GFM control structures, since it does not require any modification or constraint on the existing RPC framework but simply adjusts the slope of the nominal voltage reference.

\ To conclude this section, the conventional method and the two proposed approaches are comparatively summarized in the following Table~\ref{Comparison}.

\begin{table}[!h]
    \renewcommand{\arraystretch}{1.5} 
    \centering
    \caption{Comparison of Different Transformer Magnetization Methods}
    \label{Comparison}
    \resizebox{\columnwidth}{!}{\small   
    \begin{tabular}{|c|c|c|c|}
        \hline
         & \makecell{Conventional\\hard-magnetization} 
         & \makecell{Ultra-fast\\soft-magnetization} 
         & \makecell{Archimedean spiral\\soft-magnetization} \\
        \hline
        Flux DC offset elimination & O & O & O \\
        \hline
        Inrush current suppression & X & O & O \\
        \hline
        Surge current suppression & X & X & O \\
        \hline
        Start-up time & \makecell{Instantaneous \\ but with long \\ settling time} & $\tfrac{T_0}{2 \pi}$ & $T_0$ \\
        \hline
    \end{tabular}
    }
\end{table}

In comparison with the conventional hard-magnetization, both proposed methods eliminate the flux DC offset, thereby enabling a rapid start-up without inrush current. 
Among them, the ultra-fast soft-magnetization achieves the fastest start-up, whereas the Archimedean spiral soft-magnetization further suppresses the surge current in the presence of an LC filter, thus ensuring the most reliable black-start performance.

\section{Experimental Validation}
\label{Experiments}
To verify the effectiveness of the proposed soft-magnetization methods, experimental tests were carried out on a GFM converter connected to the grid interface transformer through an LC filter. 
The experiments compare the conventional hard-magnetization, the ultra-fast soft-magnetization, and the enhanced Archimedean spiral soft-magnetization method. 
Key performance metrics include the suppression of flux offset, mitigation of inrush and surge currents, and the overall start-up time required to complete magnetization.

\subsection{Experimental Setup}

 \begin{figure}[!t]
    \centering
    \includegraphics[width=\linewidth]{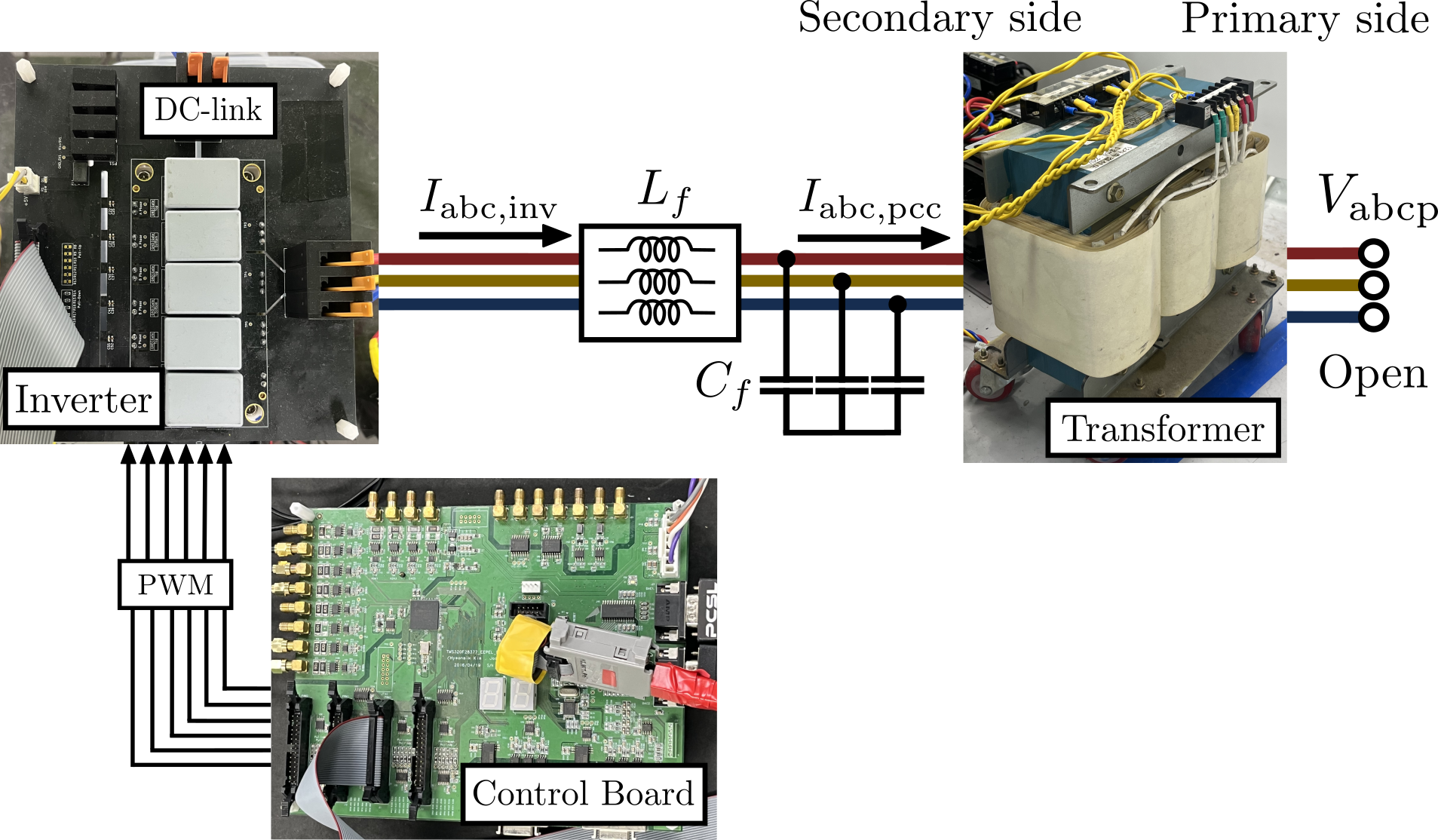}
    \caption{Experimental setup.}
    \label{Exp_setup}
\end{figure}

\begin{table}[!t]
        \centering
        \caption{Parameters of Experiment}
        \label{Exp_param}
        \resizebox{0.45\textwidth}{!}{
        \begin{tabular}{c c c}
		\hline
    \hline
    Symbol & Parameter &Nominal value\\
    \hline
        $S_{\mathrm{rated}}$ & Rated power & 5 kVA\\
        $V_{\mathrm{DC}}$ & DC-link voltage & 700 V\\
        $V_{\mathrm{rated}}$ & Converter output voltage & 400 $\mathrm{V}_{\mathrm{ll,rms}}$\\
        $I_{\mathrm{rated,peak}}$ & Rated peak current & 10.2 A\\
        $L_{\mathrm{f}}$ & Filter inductance & 3.4 mH \\
        $C_{\mathrm{f}}$ & Filter capacitance & 5 $\mu$F \\
        $f_{\mathrm{sw}}$ & Switching frequency & 8 kHz\\
        $\phi_{\mathrm{rated}}$ & Transformer rated flux & 865.6 mWb\\
        \hline
        \hline        
     \end{tabular}
        }
        
\end{table}

\ The experimental setup is shown in Fig.~\ref{Exp_setup}. A three-phase, three-limb core, Y-Y transformer was used to estimate magnetic flux-linkage. Assuming a system blackout scenario, the transformer primary side was left open, while a GFM inverter applied the voltage to the transformer secondary side. 
During this process, the flux was obtained by integrating the measured phase voltage on the primary side, and the currents at both terminals of the LC filter, namely $I_{\mathrm{inv}}$ and $I_{\mathrm{pcc}}$, were simultaneously measured.
The parameters used in the experiments are shown in Table~\ref{Exp_param}.

\begin{figure*}[!t]
    \centering
    \subfloat[]{\includegraphics[width=0.32\linewidth]{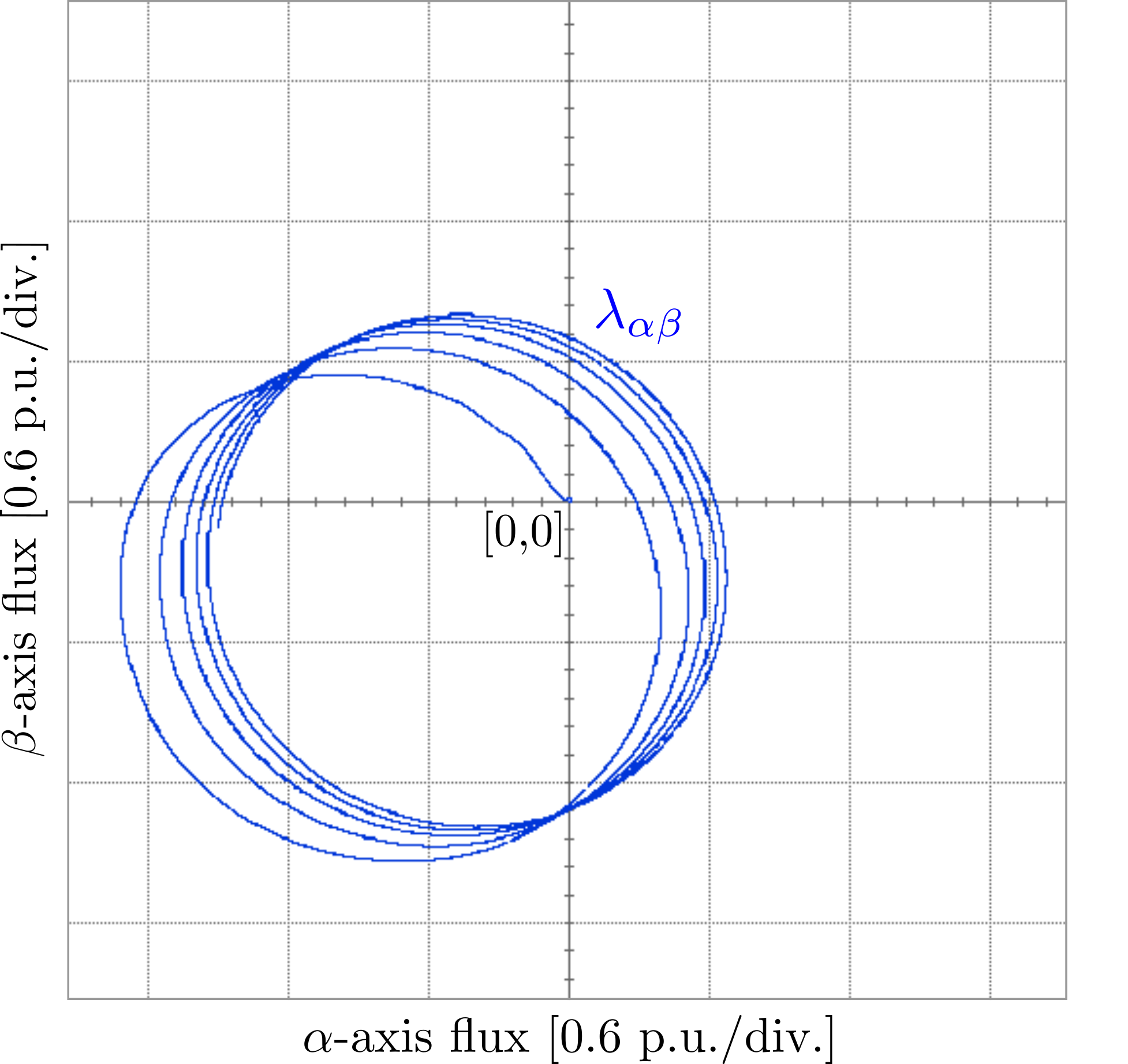}
      \label{XY_a}}    
      \hfill
    \subfloat[]{\includegraphics[width=0.32\linewidth]{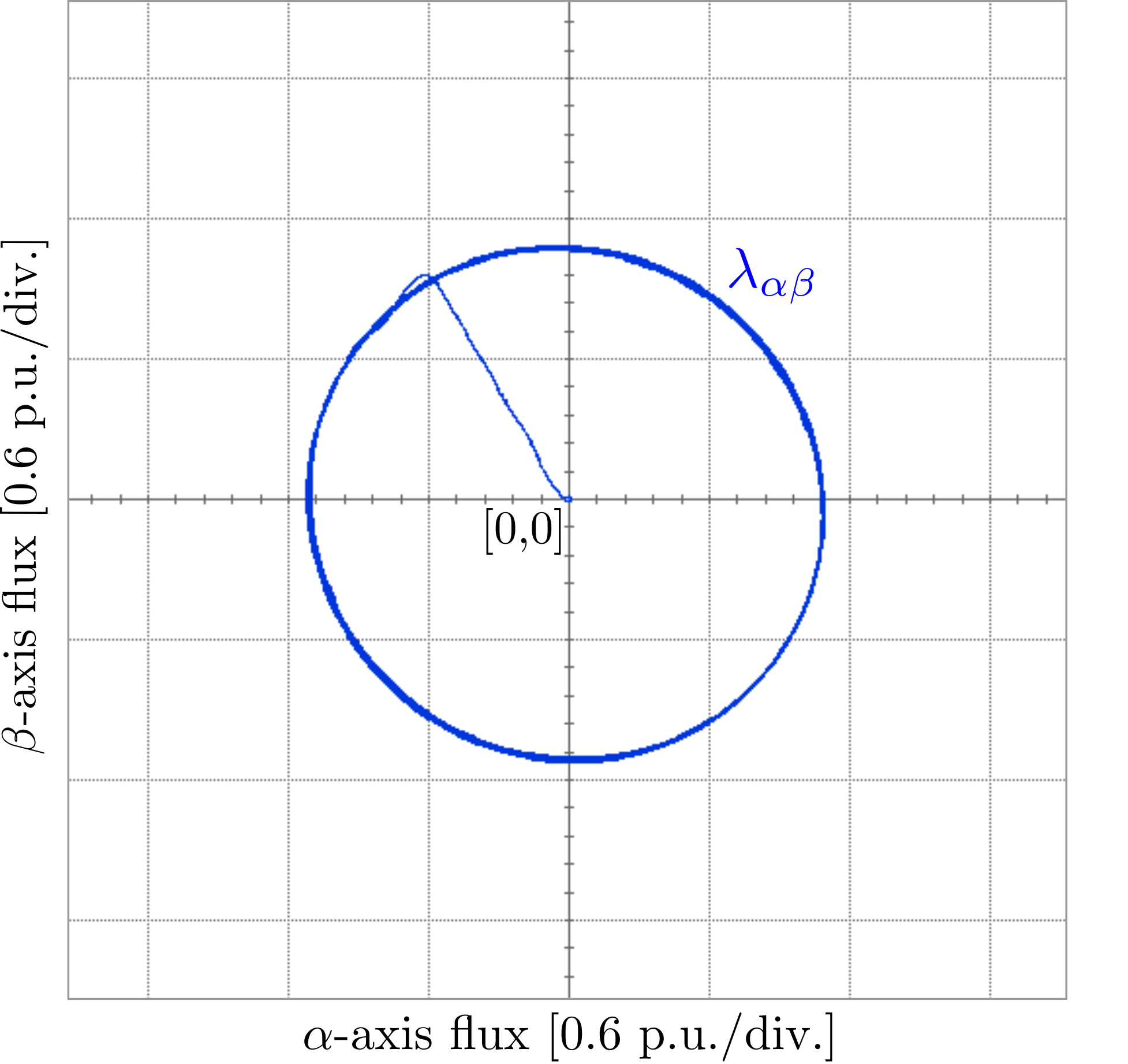}
      \label{XY_b}}
      \hfill
    \subfloat[]{\includegraphics[width=0.32\linewidth]{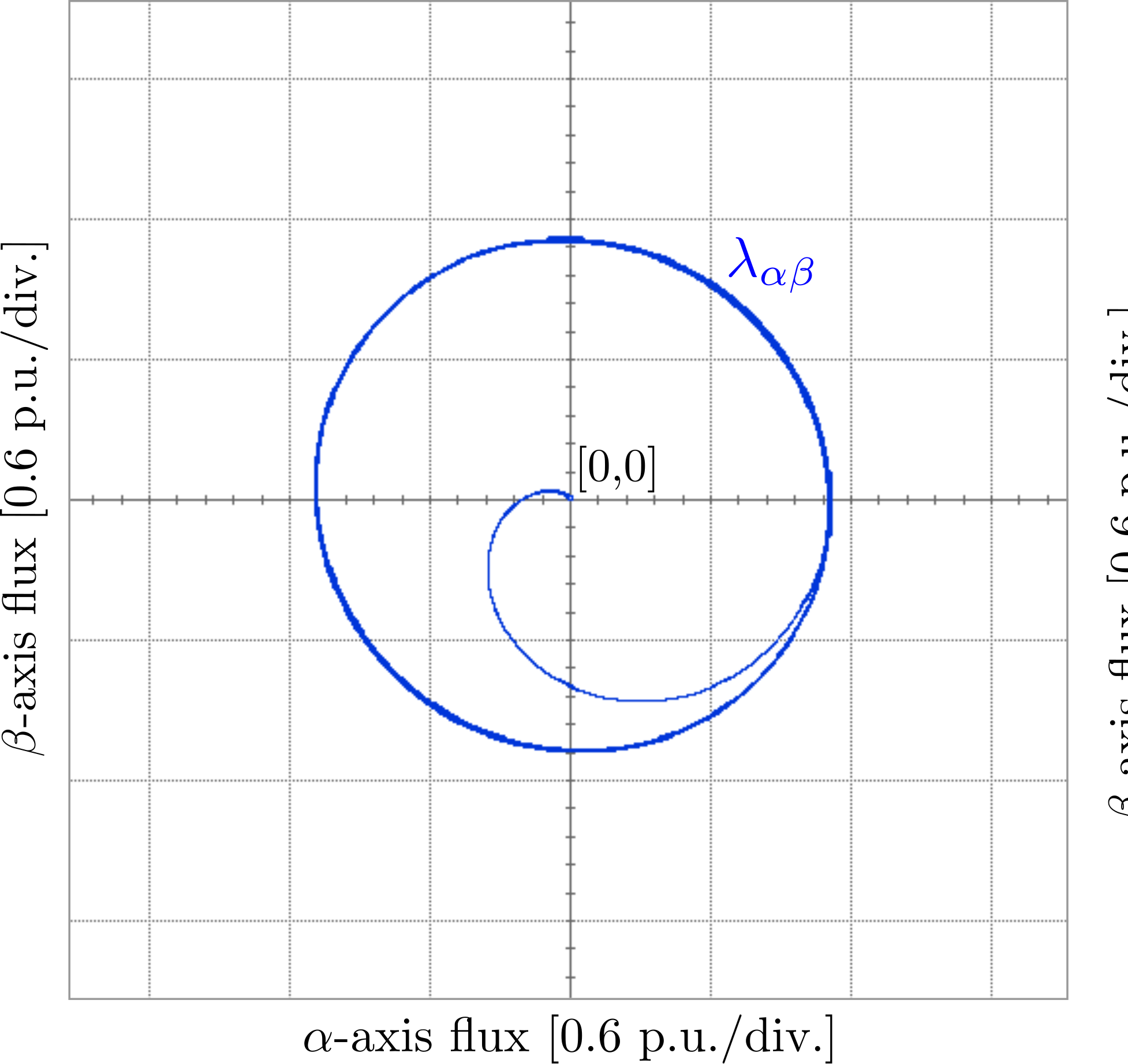}
      \label{XY_c}}
    \caption{Flux trajectories in the stationary reference frame:
    (a) conventional hard-magnetization method, 
    (b) ultra-fast soft-magnetization method, and 
    (c) enhanced Archimedean spiral soft-magnetization method.}
    \label{XY_Plot}
\end{figure*}

\begin{figure*}[!t]
    \centering
    \subfloat[]{\includegraphics[width=0.32\linewidth]{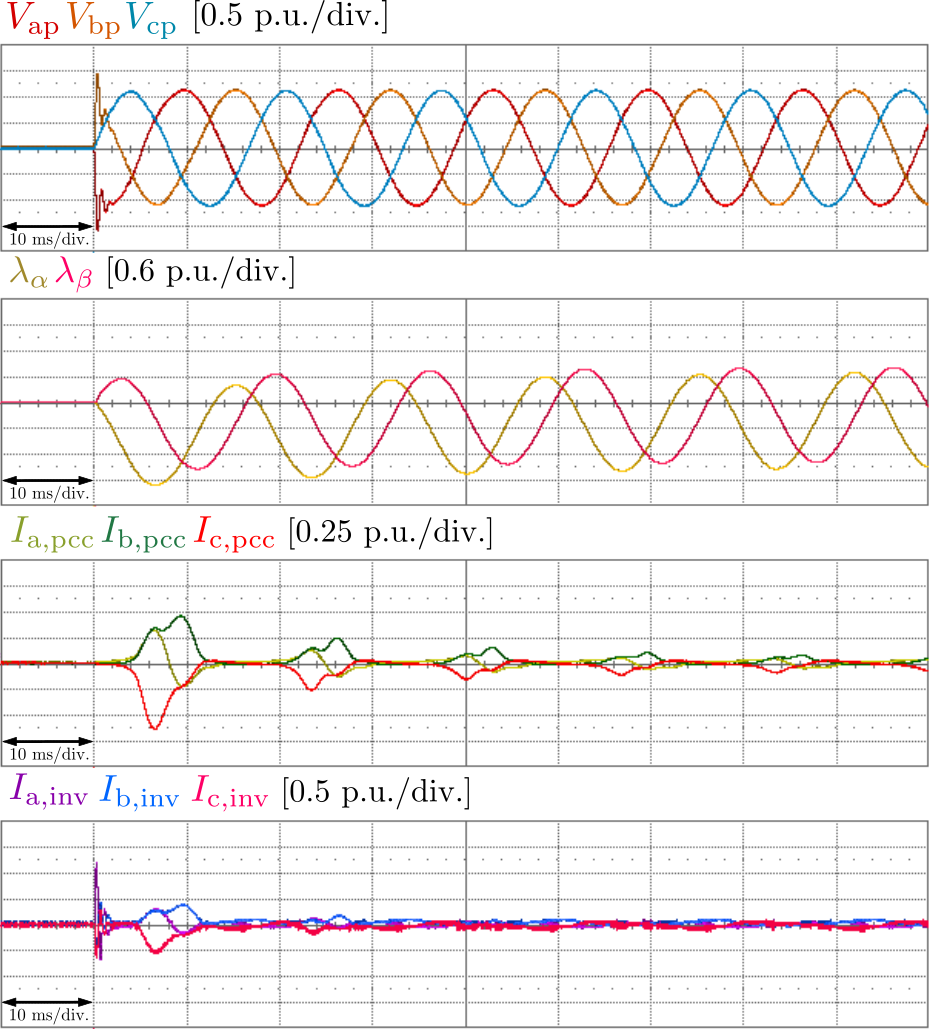}
      \label{Exp_a}}
    \hfill
    \subfloat[]{\includegraphics[width=0.32\linewidth]{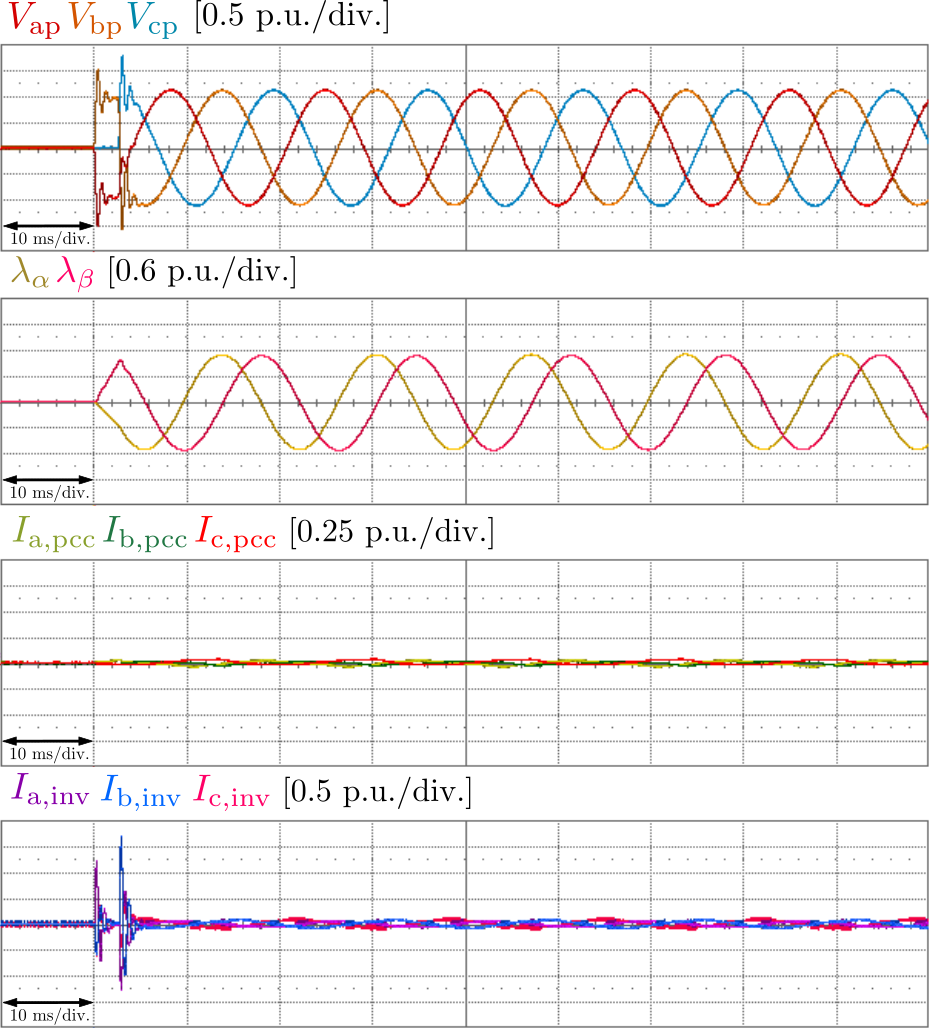}
      \label{Exp_b}}
    \hfill
    \subfloat[]{\includegraphics[width=0.32\linewidth]{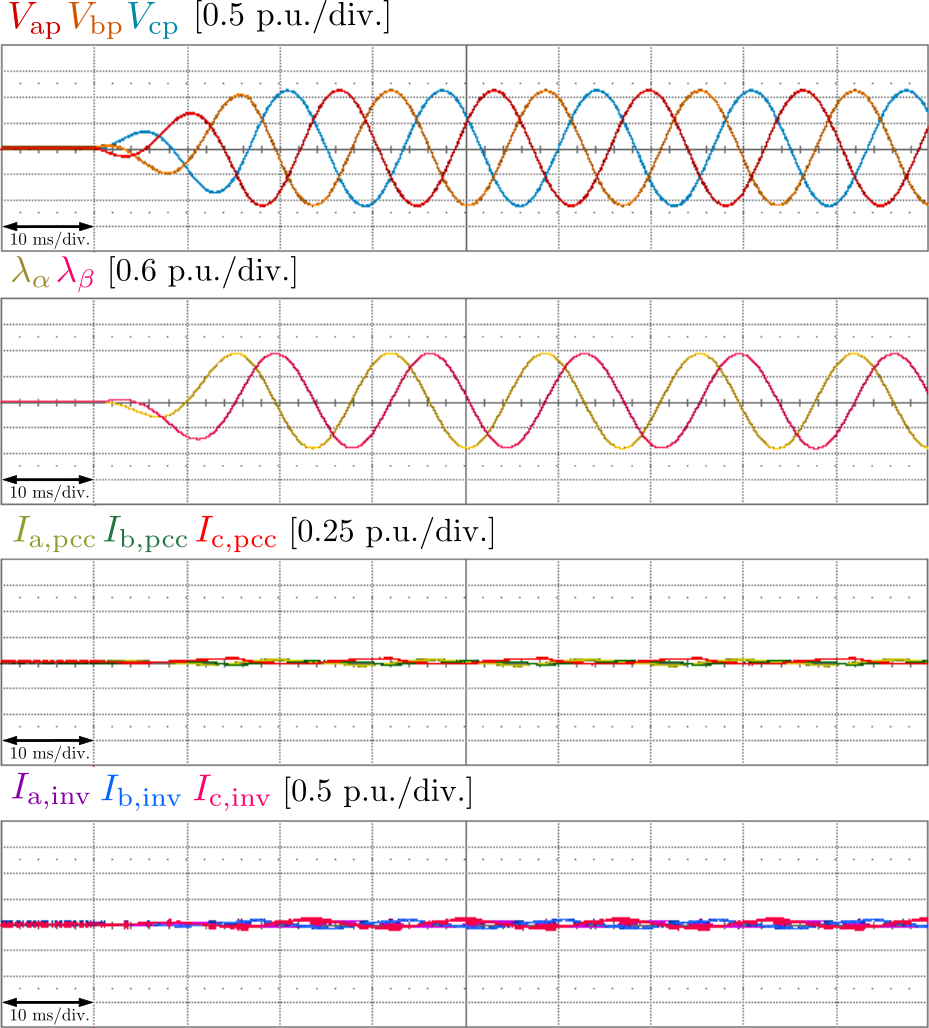}
      \label{Exp_c}}
    \caption{Experimental results of transformer magnetization without residual flux: transformer primary-side voltage, flux linkage, PCC currents, and inverter currents; 
    (a) conventional hard-magnetization method, 
    (b) ultra-fast soft-magnetization method, and 
    (c) enhanced Archimedean spiral soft-magnetization method.}
    \label{Exp_Origin}
\end{figure*}

\subsection{Experimental Results: Conventional Hard-Magnetization vs Proposed Soft-Manetization}

\ Fig.~\ref{XY_Plot} shows the trajectory of the transformer flux linkage in the stationary reference frame during start-up process. 
 Due to the phase shift within the transformer, the transformer primary-side voltage has $120^\circ$ phase difference relative to the injected voltage.
 As shown in Fig.~\ref{XY_Plot}(a), when the voltage is applied abruptly with the conventional hard-magnetization method, an offset occurs in the magnetizing flux, displacing the flux trajectory from the origin.
 Furthermore, as the DC component of the flux naturally decays with the time constant determined by the core resistance and magnetizing inductance, the flux trajectory gradually shifts toward the origin. 
 In contrast, Fig.~\ref{XY_Plot}(b) and Fig.~\ref{XY_Plot}(c) show that proposed methods effectively eliminate the flux DC offset, placing the flux circle at the origin. 
 As shown in Fig.~\ref{XY_Plot}(b), the flux initially increases and then forms a circular trajectory after being displaced by $\lambda_{0}$. 
 Fig.~\ref{XY_Plot}(c) demonstrates that, under the Archimedean spiral method, the flux follows a spiral trajectory and gradually shifts by $\lambda_{0}$ before forming a circle.

 \ This flux behavior is also evident in the time domain through the measured transformer primary-side voltage, flux linkage, PCC currents $I_{pcc}$, and LC filter currents $I_{inv}$, as shown in Fig.~\ref{Exp_Origin}. 
 During the experiment, the primary-side winding of the transformer was kept open to emulate blackout conditions, under which the magnetizing current mainly appears as the PCC current.
 As shown in Fig.~\ref{Exp_Origin}(a), when the rated voltage is applied instantaneously  with the conventional hard-magnetization method, a DC offset appears in the magnetizing flux.
 Consequently, the flux linkage acquires a DC bias, driving the core inductance into saturation and resulting in a severe inrush current.
 In contrast, Fig.~\ref{Exp_Origin}(b) and Fig.~\ref{Exp_Origin}(c) show that the proposed methods remove the DC offset of the flux linkage, thereby effectively suppressing the inrush current, and achieve very fast start-up within approximately 2.65 ms and 16.67 ms, respectively.
 However, when the ultra-fast soft-magnetization method is employed, high-frequency resonance caused by the LC filter arises at the voltage injection onset and the phase-shifting instant, as shown in Fig.~\ref{Exp_Origin}(b). 
 Thus, while this approach effectively suppresses inrush currents of the transformer at the PCC, it simultaneously induces a substantial surge current within the LC filter flowing directly through power semiconductor devices.
 In contrast, when the Archimedean spiral soft-magnetization method is applied, as shown in Fig.~\ref{Exp_Origin}(c), both the voltage magnitude and phase vary smoothly, enabling effective mitigation of both inrush current and LC filter surge current.
 As a result, the proposed method enables GFM converters to perform rapid and reliable black-start operations without the risk of converter damage caused by excessive inrush current and surge current.

\subsection{Effects of the Residual Flux}

\begin{figure*}[!t]
    \centering
    \subfloat[]{\includegraphics[width=0.32\linewidth]{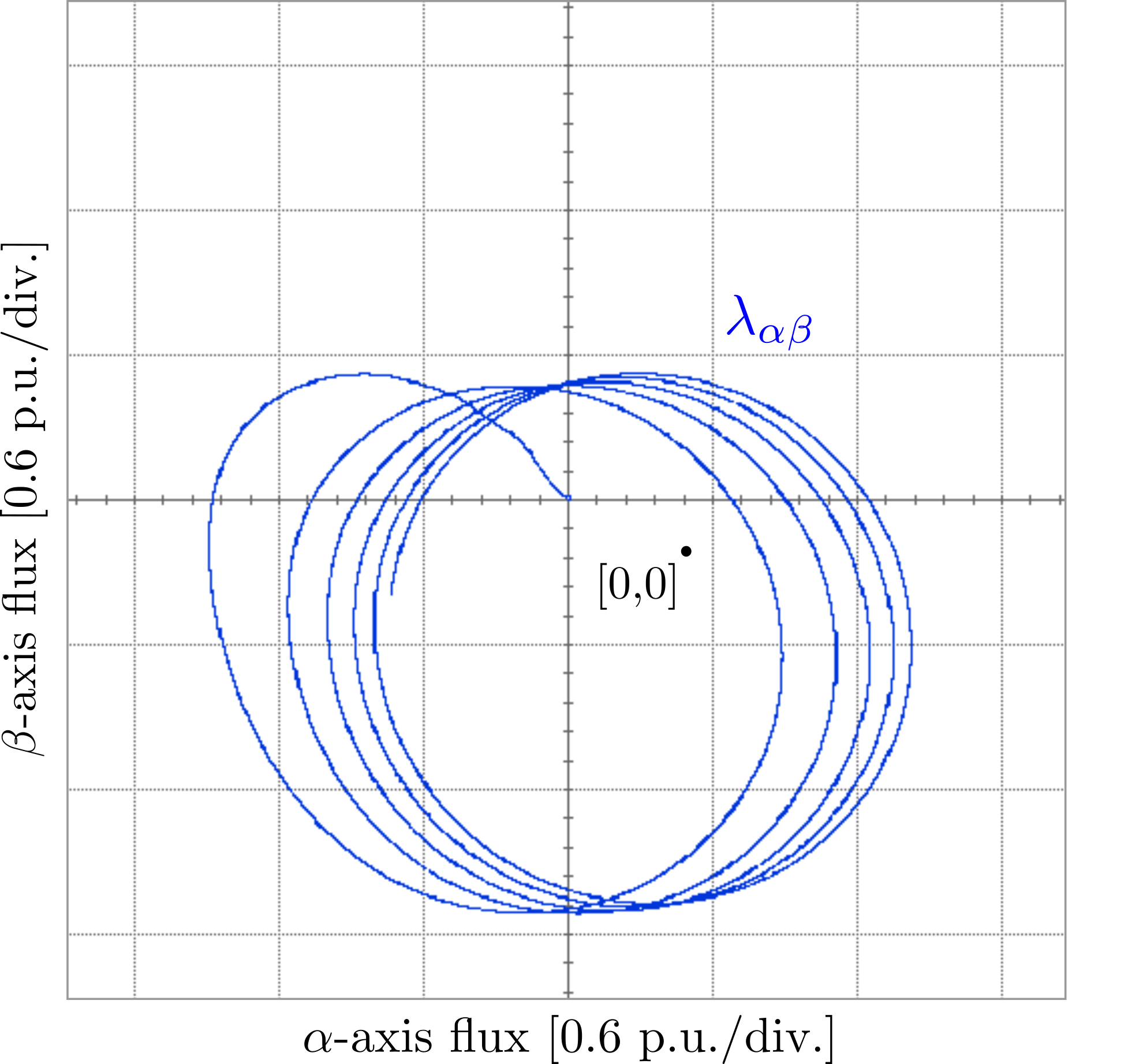}
      \label{XY_a_residual}}    
      \hfill
    \subfloat[]{\includegraphics[width=0.32\linewidth]{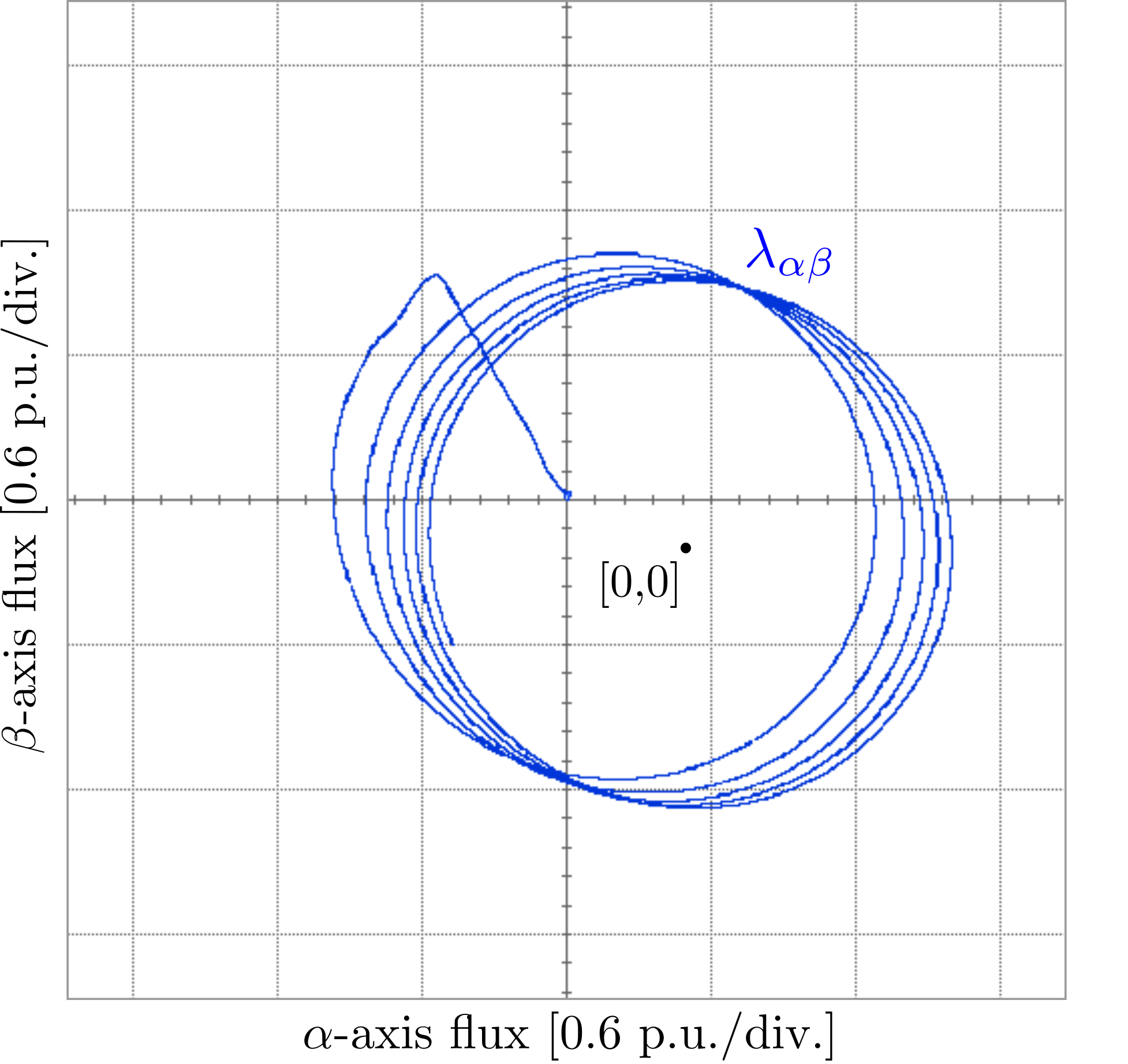}
      \label{XY_b_residual}}
      \hfill
    \subfloat[]{\includegraphics[width=0.32\linewidth]{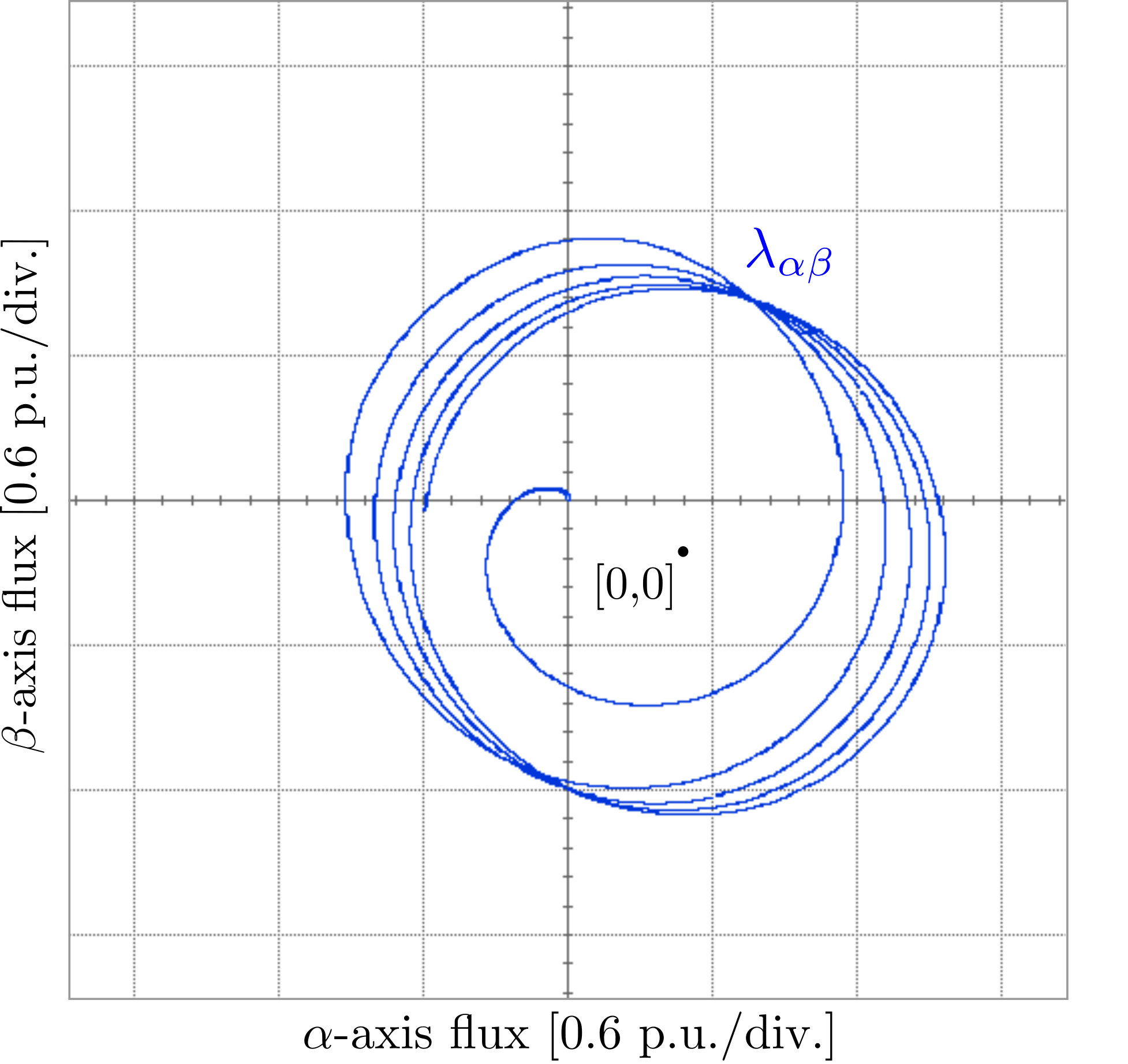}
      \label{XY_c_residual}}
    \caption{Flux trajectories in the stationary reference frame in the presence of residual flux: 
    (a) conventional hard-magnetization method, 
    (b) ultra-fast soft-magnetization method, and 
    (c) enhanced Archimedean spiral soft-magnetization method.}
    \label{XY_Plot_residual}
\end{figure*}

\begin{figure*}[!t]
    \centering
    \subfloat[]{\includegraphics[width=0.32\linewidth]{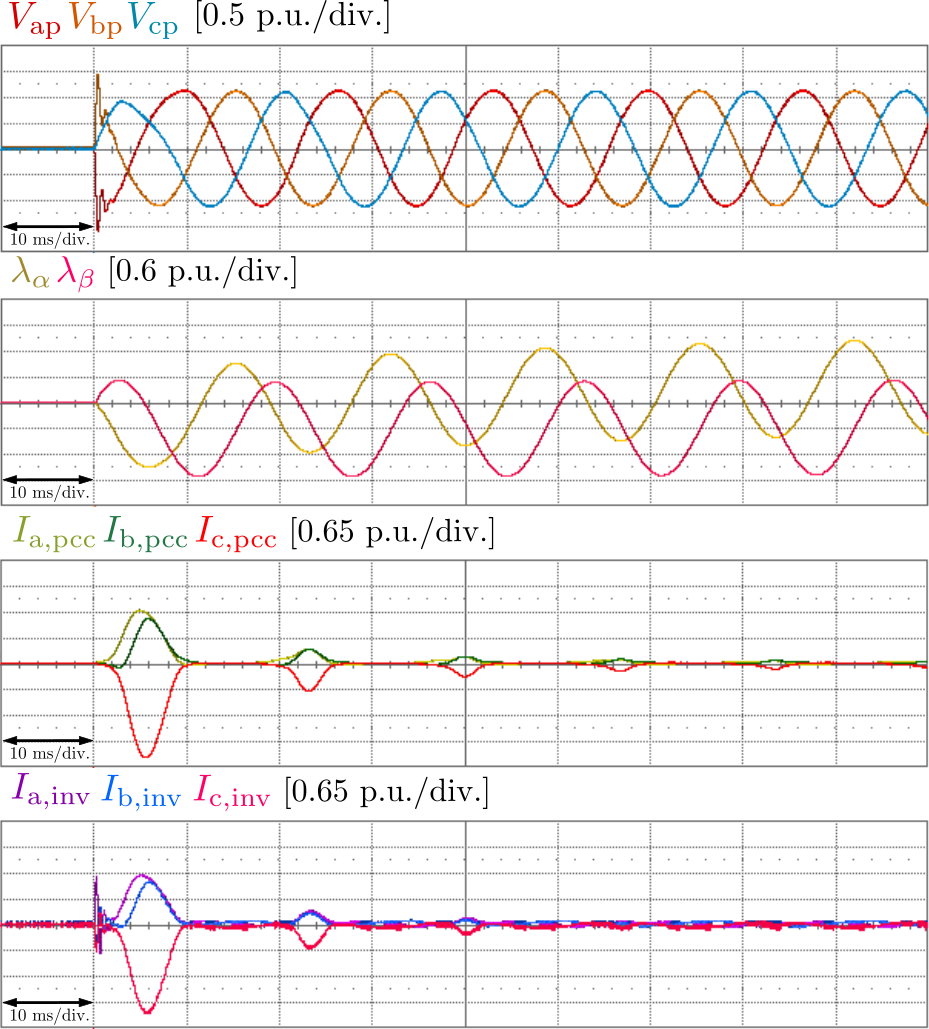}
      \label{Exp_a_residual}}
    \hfill
    \subfloat[]{\includegraphics[width=0.32\linewidth]{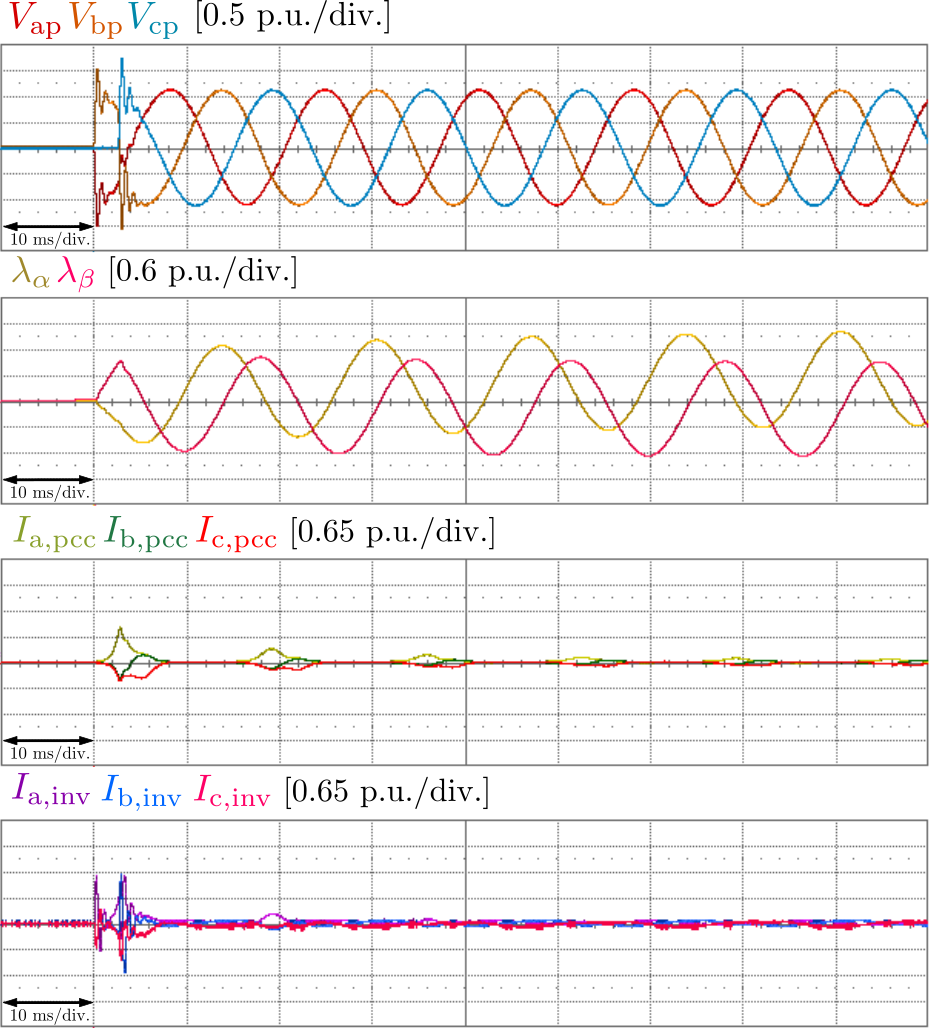}
      \label{Exp_b_residual}}
    \hfill
    \subfloat[]{\includegraphics[width=0.32\linewidth]{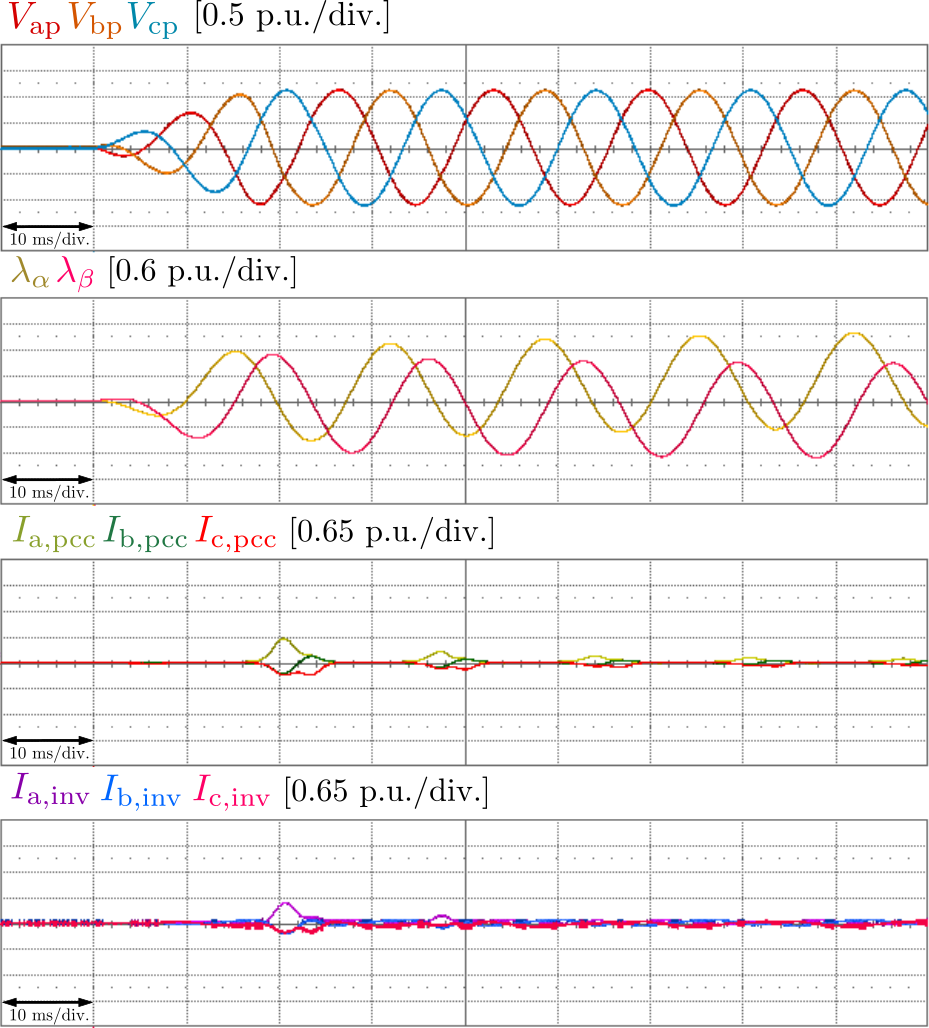}
      \label{Exp_c_residual}}
    \caption{Experimental results of transformer magnetization in the presence of residual flux: transformer primary-side voltage, flux linkage, PCC currents, and inverter currents in the presence of residual flux; 
    (a) conventional hard-magnetization method, 
    (b) ultra-fast soft-magnetization method, and 
    (c) enhanced Archimedean spiral soft-magnetization method.}
    \label{Exp_Residual}
\end{figure*}

\ The previously proposed methods assume that the initial flux trajectory starts at the origin, i.e., the transformer core retains no residual flux after a blackout.
 Under this assumption, the energization process can achieve a rapid start-up without inrush current within the desired time frame. However, in practical blackout scenarios, when the inverter is suddenly switched off and the applied voltage across the transformer instantaneously drops to zero, 
 the core hysteresis may cause residual flux to remain in the transformer core.
 When residual flux remains in the transformer core, the experimental results obtained by applying the same start-up methods are shown in Fig.~\ref{XY_Plot_residual} and Fig.~\ref{Exp_Residual}.
 In order to establish residual flux in the transformer core, DC voltages of (10~V, 0~V, -10~V) were applied to the three phases for a fixed duration to intentionally drive the core into the same level of saturation for fair comparison, after which application of DC voltages was terminated.
 As shown in Fig.~\ref{XY_Plot_residual}, when the start-up is carried out in the presence of residual flux, the initial point of the flux trajectory no longer coincides with the origin $(0,0)$.
 In this case, since the transformer primary-side voltage was directly integrated with time, the trajectory appears to originate from the origin in Fig.~\ref{XY_Plot_residual}. 
 However, the actual origin of the flux trajectory corresponds to the location of the residual flux, which is represented as the new reference point in the scope.
 Therefore, as shown in Fig.~\ref{Exp_Residual}(a), when hard-magnetization is applied under the worst-case condition with residual flux, an inrush current exceeding 2~p.u. is produced, which can cause severe damage to the inverter.
 When the proposed soft-magnetization methods are applied, the flux DC offset is relatively reduced, as shown in Fig.~\ref{XY_Plot_residual}(b) and Fig.~\ref{XY_Plot_residual}(c), but cannot be completely eliminated. 
 Consequently, as illustrated in Fig.~\ref{Exp_Residual}(b) and Fig.~\ref{Exp_Residual}(c), inrush currents still occur with magnitudes close to 1~p.u.
 This represents a substantial level compared with Fig.~\ref{Exp_Origin}(b) and Fig.~\ref{Exp_Origin}(c), indicating that residual flux makes it difficult to precisely shape the flux trajectory as intended.
 Therefore, prior to applying the proposed start-up methods, demagnetization must be performed to reset the flux trajectory to the origin to ensure a reliable and fast black-start process.

\subsection{Demagnetization Process before Applying Soft-Magnetization Methods}

 \begin{figure}[!t]
    \centering
    \subfloat[]{\includegraphics[width= 0.6\linewidth]{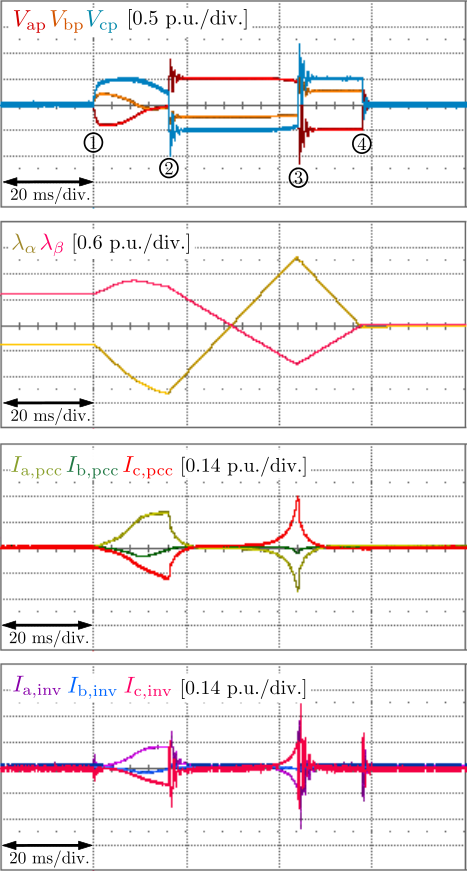}
      \label{Demag_exp}}
    \hfill
    \subfloat[]{\includegraphics[width=0.67\linewidth]{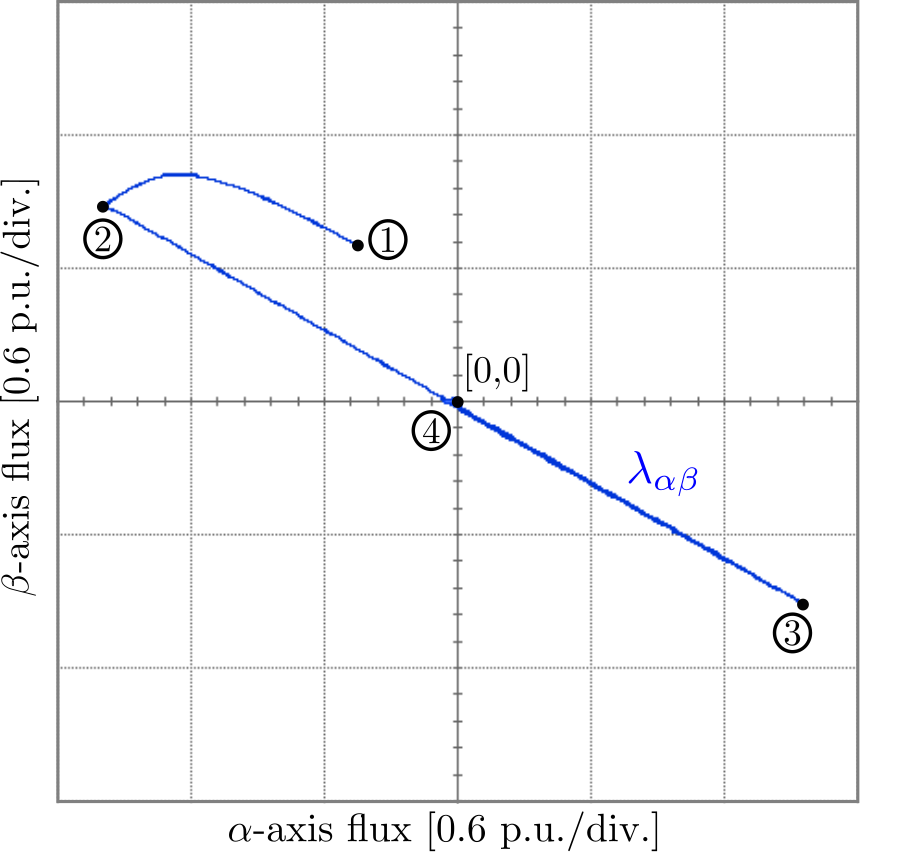}
      \label{Demag_XY}}
    \caption{Demagnetization process: (a) transformer primary-side voltage, flux linkage, PCC currents, and inverter currents in the presence of residual flux, (b) flux trajectories in the stationary reference frame.}
    \label{Demag_process}
\end{figure}

In this paper, the inherent voltage profiling capability of inverters with high bandwidth is exploited to perform transformer demagnetization during the initial start-up stage.
The demagnetization process of the transformer using the inverter was experimentally validated, as shown in Fig.~\ref{Demag_process}.

\begin{figure}[!t]
    \centering
    \subfloat[]{\includegraphics[width=0.8\linewidth]{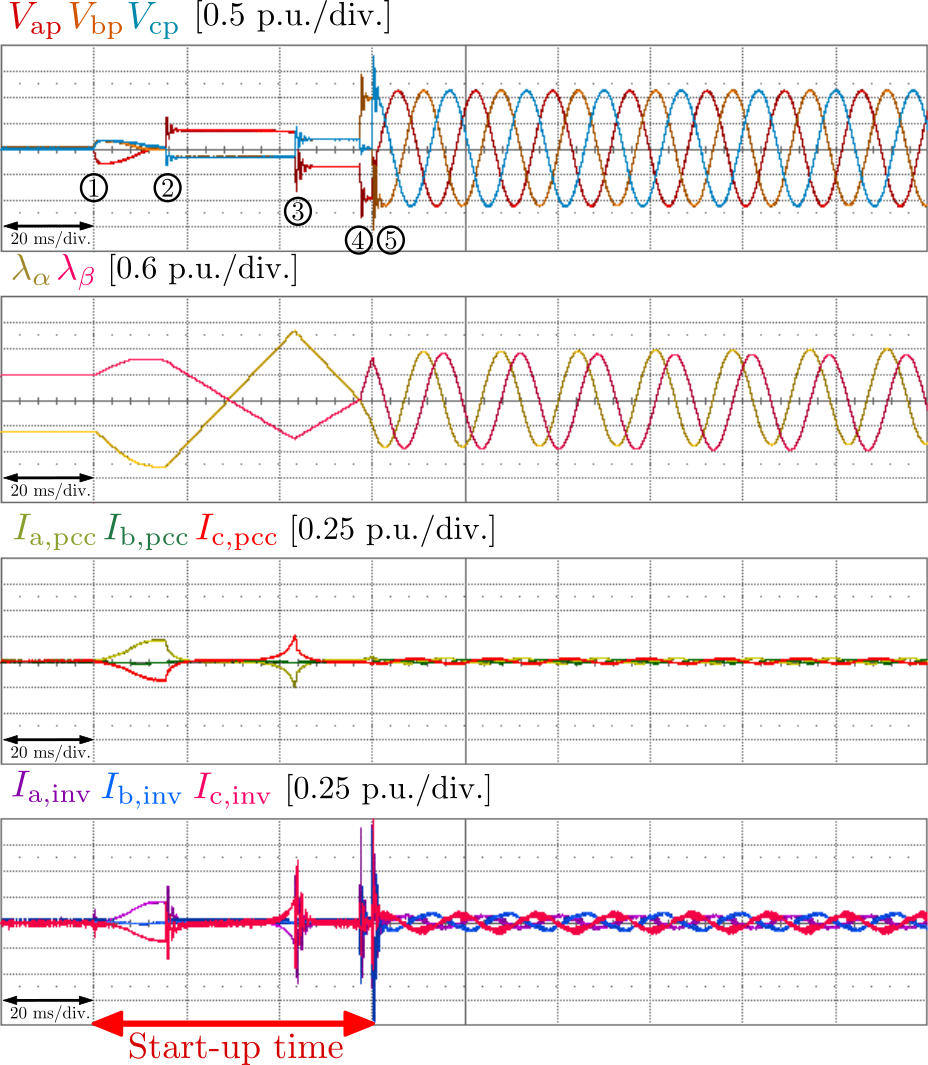}
      \label{Exp_total_PEDG}}
    \hfill
    \subfloat[]{\includegraphics[width=0.7\linewidth]{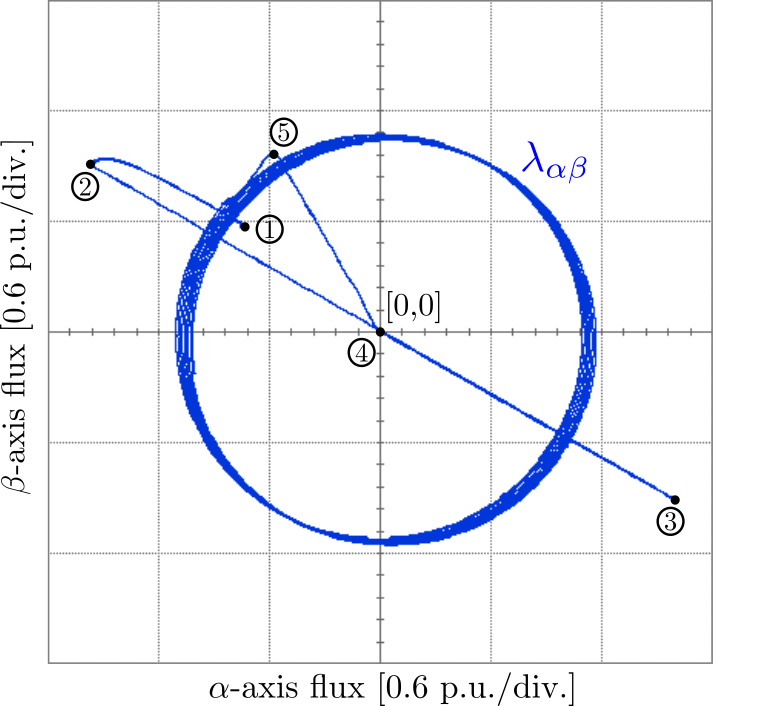}
      \label{XY_total_PEDG}}
    \caption{Experimental results of the total start-up scenario of the ultra-fast soft-magnetization method in the presence of residual flux: 
    (a) transformer flux linkage, PCC currents, inverter currents, and applied voltage, 
    (b) transformer flux linkage in the stationary reference frame.}
    \label{Total_PEDG}
\end{figure}

\begin{figure}[!t]
    \centering
    \subfloat[]{\includegraphics[width=0.8\linewidth]{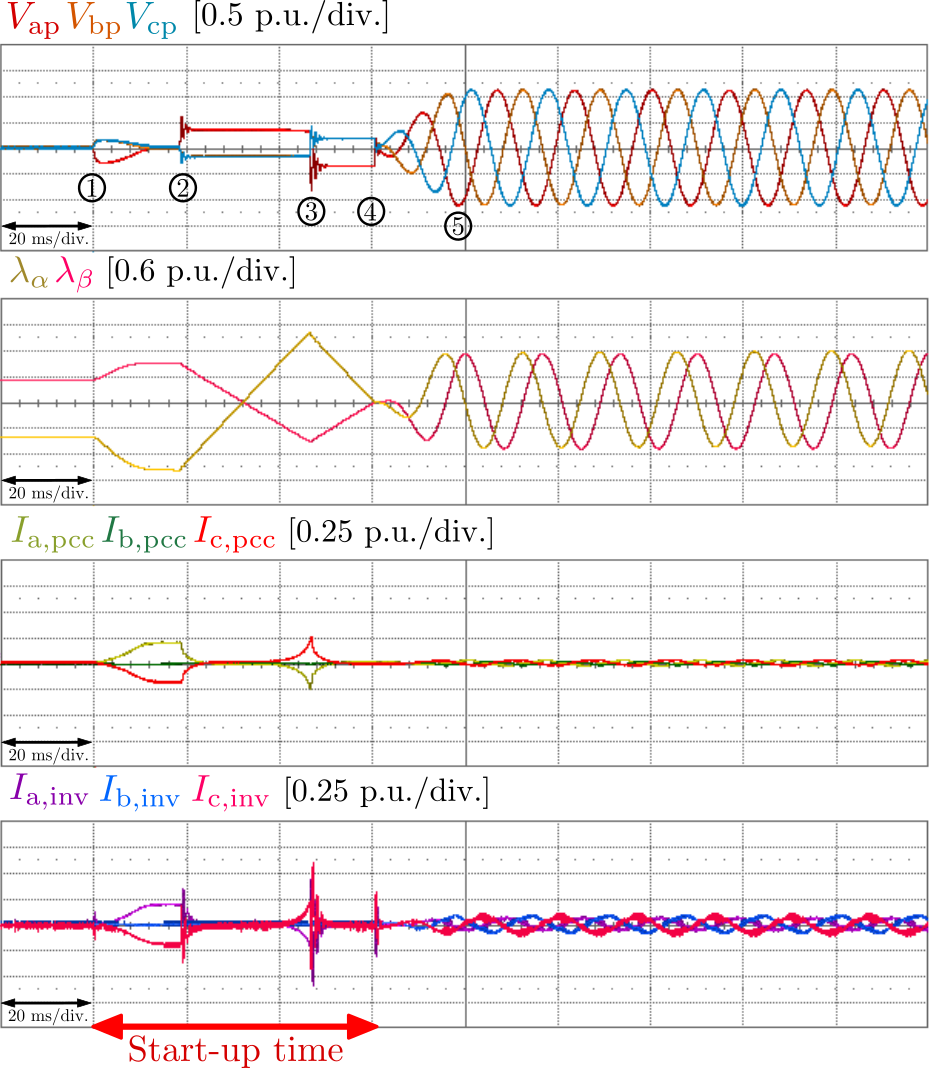}
      \label{Exp_total_Snail}}
    \hfill
    \subfloat[]{\includegraphics[width=0.7\linewidth]{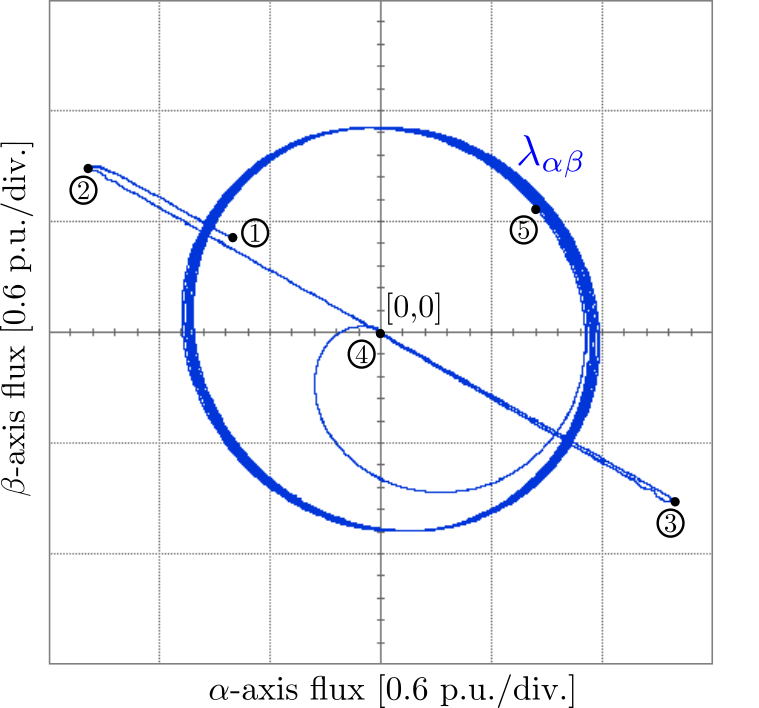}
      \label{XY_total_Snail}}
    \caption{Experimental results of the total start-up scenario of the enhanced archimedean spiral soft-magnetization method in the presence of residual flux: 
    (a) transformer flux linkage, PCC currents, inverter currents, and applied voltage, 
    (b) transformer flux linkage in the stationary reference frame.}
    \label{Total_Snail}
\end{figure}

The sequence of the demagnetization process is as follows based on the method developed in \cite{Demagnetization}:
\begin{enumerate}
  \item Since the residual flux in the transformer core is unknown, the core is first driven into pre-defined deep saturation through inverter current control. 
  To ensure that the core fully enters the saturation region, three phase current references of $(3~\mathrm{A}, 0~\mathrm{A}, -3~\mathrm{A})$ are applied. 
  At this point, considering the rated current and no-load current of the transformer, the current threshold was appropriately set to ensure that the core enters the saturation region.
  During this process, the current response is monitored until the saturation region is reached and the abrupt rise of inrush current is suppressed due to closed-loop current control.
  \item When the $a$ phase current reaches $3~\mathrm{A}$, a DC voltage of $(-10~\mathrm{V}, 0~\mathrm{V}, 10~\mathrm{V})$ is applied to drive the core into saturation in the opposite direction.
   Once the $a$ phase current decreases to $-3~\mathrm{A}$, the voltage application is terminated.
    During this process, the elapsed time from the beginning to the end of voltage injection is measured.
  \item By applying the DC voltage in the opposite direction again for half of the previously measured duration, the core flux increases in the reverse direction and returns to the origin.
   In this way, the transformer can be fully demagnetized even in the presence of unknown residual flux and ready for the subsequent start-up process proposed in this paper.
\end{enumerate}

As shown in Fig.~\ref{Demag_process}(b), the initial residual flux is located at point 1.
Through the first step, the fluxes in phase $a$ and $c$ are saturated in the positive and negative directions, respectively.
When the residual flux reaches point 2, and the phase $a$ current reaches 3 A as shown in Fig.~\ref{Demag_process}(a), the polarity of the applied voltage is reversed, driving the flux into saturation in the opposite direction, thereby reaching point 3.
During this process, the duration of voltage application is measured. 
By applying the voltage in the reverse direction again for half of this measured time, the flux finally returns to the origin, as point 4.

\ Therefore, by incorporating the demagnetization sequence prior to the start-up, the proposed soft-magnetization based black-start can be reliably executed, as shown in Fig.~\ref{Total_PEDG} and \ref{Total_Snail}.
Fig.~\ref{Total_PEDG} shows the total start-up process of the ultra-fast soft-magnetization method and Fig.~\ref{Total_Snail} shows the Archimedean spiral soft-magnetization method. 
As shown in Fig.~\ref{Total_PEDG}(b) and Fig.~\ref{Total_Snail}(b), when residual flux exists initially (point 1), the demagnetization process relocates the flux trajectory to the origin (point 2--point 4). 
Subsequently, by applying the soft-magnetization method, the voltage is ramped up and the flux begins to rotate perfectly around the origin without any DC offset (point 5).
As shown in Fig.~\ref{Total_PEDG}(a) and Fig.~\ref{Total_Snail}(a), in comparison with Fig.~\ref{Exp_Residual}(b) and Fig.~\ref{Exp_Residual}(c), it can be observed that even under the same residual flux condition, the demagnetization process effectively suppresses the inrush current during start-up.
Moreover, even with the inclusion of the demagnetization process, the total start-up time remains on the order of a few milliseconds, approximately 60 ms. 
Therefore, by accurately designing the flux trajectory and profiling the applied voltage, a highly reliable and inrush-current-free black-start has been successfully achieved.



\IEEEpubidadjcol

\section{Conclusion}
\label{conclusion}
This paper presented two novel transformer soft-magnetization methods for enabling rapid and reliable black-start of grid-forming converters.  
First, by interpreting the complex three-phase voltages and fluxes in the stationary reference frame, an intuitive analysis of three-phase transformer core saturation was provided.  
Based on this framework, the flux trajectory was deliberately designed to eliminate DC offset, forming the foundation of the proposed methods.  
The ultra-fast soft-magnetization method eliminates flux DC offset within a fraction of the fundamental cycle, achieving the fastest start-up.  
Meanwhile, the enhanced Archimedean spiral soft-magnetization method ensures smooth voltage magnitude and phase transitions, thereby suppressing both inrush current and LC filter associated surge current.  
Additionally, a practical demagnetization process was introduced and experimentally validated to address residual flux conditions, further enhancing reliability in practice.  
The combined results demonstrate that accurate flux trajectory design and voltage profiling enable black-start within only a few milliseconds, offering a robust and semiconductor-safe solution for next-generation inverter-based power grids.




 

\phantomsection
\bibliographystyle{IEEEtran}
\bibliography{JESTPE_Reference.bib}

@ARTICLE{GFM_Contol_Basic,
  author={Rocabert, Joan and Luna, Alvaro and Blaabjerg, Frede and Rodríguez, Pedro},
  journal={IEEE Transactions on Power Electronics}, 
  title={Control of {Power} {Converters} in {A}{C} {Microgrids}}, 
  year={2012},
  volume={27},
  number={11},
  pages={4734-4749},
  doi={10.1109/TPEL.2012.2199334}}

@article{rosso_grid-forming_2021,
	title = {Grid-{Forming} {Converters}: {Control} {Approaches}, {Grid}-{Synchronization}, and {Future} {Trends}—{A} {Review}},
	volume = {2},
	issn = {2644-1241},
	shorttitle = {Grid-{Forming} {Converters}},
	doi = {10.1109/OJIA.2021.3074028},
	urldate = {2025-02-26},
	journal = {IEEE Open Journal of Industry Applications},
	author = {Rosso, Roberto and Wang, Xiongfei and Liserre, Marco and Lu, Xiaonan and Engelken, Soenke},
	year = {2021},
	pages = {93--109},
}

@article{GFM_IBR,
author = {Khan, Musa and Wu, Wenchuan and Li, Li},
title = {Grid-{forming} {control} for {inverter}-{based} {resources} in {power} {systems}: {A} {review} on its {operation}, {system} {stability}, and {prospective}},
journal = {IET Renewable Power Generation},
volume = {18},
number = {6},
pages = {887-907},
doi = {https://doi.org/10.1049/rpg2.12991},
year = {2024}
}

@inproceedings{alassi_performance_2020,
	title = {Performance {Evaluation} of {Four} {Grid}-{Forming} {Control} {Techniques} with {Soft} {Black}-{Start} {Capabilities}},
	doi = {10.1109/ICRERA49962.2020.9242758},
	urldate = {2025-02-26},
	booktitle = {2020 9th {International} {Conference} on {Renewable} {Energy} {Research} and {Application} ({ICRERA})},
	author = {Alassi, Abdulrahman and Ahmed, Khaled and Egea-Alvarez, Agusti and Ellabban, Omar},
	month = sep,
	year = {2020},
	note = {ISSN: 2572-6013},
	pages = {221--226},
}

@INPROCEEDINGS{Black_cap_GFM,
  author={Sawant, Jay and Haldar, Arundhuti and Jain, Rishabh and Luck, Ben},
  booktitle={IEEE Energy Conversion Congress and Exposition (ECCE)}, 
  title={Black-{Starting} {Microgrids} Using {Inverter}-{Based} {Distributed} {Energy} {Resources}}, 
  year={2024},
  volume={},
  number={},
  pages={939-946},
  doi={10.1109/ECCE55643.2024.10861204}}

@INPROCEEDINGS{Black_start_cap2,
  author={Xie, P. and Liu, Y. and Lin, X. and Lu, Q. and Xue, Y. and Zhang, Z. and Xu, Z.},
  booktitle={Annual Meeting of CSEE Study Committee of HVDC and Power Electronics (HVDC 2023)}, 
  title={{Black} {Start} {Strategy} {of} {Grid-Forming} {CIGs} {Supplying} {Passive} {Island} {Grids}},  
  year={2023},
  volume={2023},
  number={},
  pages={1-5},
  doi={10.1049/icp.2023.3019}}

@inproceedings{burroughs_black_2023,
	address = {Orlando, FL, USA},
	title = {Black {Start} with {Inverter}-{Based} {Resources}: {Hardware} {Testing}},
	copyright = {https://doi.org/10.15223/policy-029},
	isbn = {978-1-6654-6441-3},
	shorttitle = {Black {Start} with {Inverter}-{Based} {Resources}},
	doi = {10.1109/PESGM52003.2023.10252637},
	language = {en},
	urldate = {2025-02-27},
	booktitle = {2023 {IEEE} {Power} \& {Energy} {Society} {General} {Meeting} ({PESGM})},
	publisher = {IEEE},
	author = {Burroughs, Hannah and Klauber, Cecilia and Sun, Chih-Che and Culler, Megan},
	month = jul,
	year = {2023},
	pages = {1--5},
}

@ARTICLE{NRE,
  author={Zhan, Changjiang and Ge, Jing and Hao, Fei and Zhu, Haobing and Wang, Chen and Wang, Nannan and Wu, Heng and Wang, Xiongfei},
  journal={IEEE Power Electronics Magazine}, 
  title={{Assuring} {Security} {and} {Stability} {of} {a} {Remote/Islanded} {Large} {Electric} {Power} {System} {With} {High} {Penetration} {of} {Variable} {Renewable} {Energy} {Resources}}, 
  year={2025},
  volume={12},
  number={1},
  pages={53-63},
  doi={10.1109/MPEL.2025.3527779}}

@ARTICLE{CS_Method,
  author={Ni, Haimiao and Fang, Shuhua and Lin, Heyun},
  journal={IEEE Transactions on Power Delivery}, 
  title={{A} {Simplified} {Phase-Controlled} {Switching} {Strategy} {for} {Inrush} {Current} {Reduction}}, 
  year={2021},
  volume={36},
  number={1},
  pages={215-222},
  doi={10.1109/TPWRD.2020.2984234}}

@INPROCEEDINGS{Prefluxing_VR,
  author={Zhu, Jinli and Abdelhadi, Ahmad Fares and Li, Yuan and Peng, Fang Z.},
  booktitle={2024 IEEE Energy Conversion Congress and Exposition (ECCE)}, 
  title={{Instant} {Transformer} {Energization} {for} {Inverter} {Based} {Resources} {Based} {on} {Pre-fluxing} {and} {Virtual} {Resistance} {Control}}, 
  year={2024},
  volume={},
  number={},
  pages={849-854},
  doi={10.1109/ECCE55643.2024.10861563}}

@article{PIR,
  title={{Blackstart} {from} {HVDC-connected} {Offshore} {Wind}: {Hard} {Versus} {Soft} {Energization}},
  author={Jain, Anubhav and Sabor{\'\i}o-Romano, Oscar and Sakamuri, Jayachandra N and Cutululis, Nicolaos A},
  journal={IET renewable power generation},
  volume={15},
  number={1},
  pages={127--138},
  year={2021},
  publisher={Wiley Online Library}
}

@ARTICLE{Soft_Energization2,
  author={Alassi, Abdulrahman and Ahmed, Khaled H. and Egea-Alvarez, Agusti and Foote, Colin},
  journal={IEEE Transactions on Power Delivery}, 
  title={{Transformer} {Inrush} {Current} {Mitigation} {Techniques} {for} {Grid-Forming} {Inverters} {Dominated} {Grids}}, 
  year={2023},
  volume={38},
  number={3},
  pages={1610-1620},
  doi={10.1109/TPWRD.2022.3218923}}

@INPROCEEDINGS{Ramp_Time_Estimation,
  author={Alassi, Abdulrahman and Ahmed, Khaled and Egea-Alvarez, Agusti and Foote, Colin},
  booktitle={2021 56th International Universities Power Engineering Conference (UPEC)}, 
  title={{Soft} {Transformer} {Energization}: {Ramping} {Time} {Estimation} {Method} {for} {Inrush} {Current} {Mitigation}}, 
  year={2021},
  volume={},
  number={},
  pages={1-6},
  doi={10.1109/UPEC50034.2021.9548203}}

@INPROCEEDINGS{SPC,
  author={Rodriguez, P. and Candela, I. and Luna, A.},
  booktitle={2013 IEEE Energy Conversion Congress and Exposition}, 
  title={{Control} {of} {PV} {Generation} {Systems} {Using} {the} {Synchronous} {Power} {Controller}}, 
  year={2013},
  volume={},
  number={},
  pages={993-998},
  doi={10.1109/ECCE.2013.6646811}}

@ARTICLE{PSC,
  author={Zhang, Lidong and Harnefors, Lennart and Nee, Hans-Peter},
  journal={IEEE Transactions on Power Systems}, 
  title={{Power-Synchronization} {Control} {of} {Grid-Connected} {Voltage-Source} {Converters}}, 
  year={2010},
  volume={25},
  number={2},
  pages={809-820},
  doi={10.1109/TPWRS.2009.2032231}}

@ARTICLE{Flux_Volt_JH1,
  author={Yun, Jonghun and Cui, Shenghui},
  journal={IEEE Transactions on Power Electronics}, 
  title={{Enhanced} {Active} {Thermal} {Balancing} {Strategy} {for} {Three-Phase} {Dual-Active} {Bridge} {Converter} {Suppressing} {Transformer} {Flux} {Saturation}}, 
  year={2025},
  volume={40},
  number={4},
  pages={5014-5024},
  doi={10.1109/TPEL.2024.3517534}}

@ARTICLE{Flux_Volt_JH2,
  author={Yun, Jonghun and Cui, Shenghui and Sul, Seung-Ki},
  journal={IEEE Transactions on Power Electronics}, 
  title={{Instantaneous} {Pulse} {Pattern} {Control} {for} {Optimized} {Dynamic} {Performance} {of} {Three-Phase} {Dual-Active} {Bridge} {Converter}}, 
  year={2025},
  volume={40},
  number={7},
  pages={9019-9033},
  doi={10.1109/TPEL.2025.3543493}}

@ARTICLE{Flux_Volt_JH3,
  author={Yun, Jonghun and Yoo, Jiwon and Cui, Shenghui and Sul, Seung-Ki},
  journal={IEEE Transactions on Power Electronics}, 
  title={{Model} {Predictive} {Control} {for} {Six-Step} {Operation} {of} {PMSM} {Based} {on} {Adapted} {Fast} {Gradient} {Method}}, 
  year={2023},
  volume={38},
  number={5},
  pages={5952-5962},
  doi={10.1109/TPEL.2023.3240510}}

@INPROCEEDINGS{PEDG,
  author={Lee, Jiyu and Yun, Jonghun and Lee, Jaekeun and Wu, Heng and Jung, Jae-Jung and Cui, Shenghui},
  booktitle={2025 IEEE 16th International Symposium on Power Electronics for Distributed Generation Systems (PEDG)}, 
  title={{Ultra-Fast} {Black-Start} {Method} {of} {the} {Grid-Forming} {Converter} {for} {Electronic} {Power} {Grids} {with} {Transformer} {Soft} {Magnetization}}, 
  year={2025},
  volume={},
  number={},
  pages={562-566},
  doi={10.1109/PEDG62294.2025.11060300}}

@ARTICLE{Demagnetization,
  author={de León, Francisco and Farazmand, Ashkan and Jazebi, Saeed and Deswal, Digvijay and Levi, Raka},
  journal={IEEE Transactions on Power Delivery}, 
  title={{Elimination} {of} {Residual} {Flux} {in} {Transformers} {by} {the} {Application} {of} {an} {Alternating} {Polarity} {DC} {Voltage} {Source}}, 
  year={2015},
  volume={30},
  number={4},
  pages={1727-1734},
  doi={10.1109/TPWRD.2014.2377199}}

\newpage

\vfill

\end{document}